\documentclass[twocolumn, secnumarabic, amssymb, nobibnotes, pre, superscriptaddress]{revtex4-2}

\usepackage{graphicx} 
\usepackage{amsmath}
\usepackage{float}
\usepackage{color}
\usepackage[english]{babel}
\usepackage{array}
\usepackage[T1]{fontenc}
\usepackage[usenames,dvipsnames]{xcolor}
\usepackage{setspace}
\usepackage{hyperref}
\usepackage{bm}
\usepackage{lipsum} 
\usepackage{tikz}
\usepackage{stmaryrd}
\usepackage{wasysym}
\usepackage{blkarray}
\usepackage{mathtools}
\usepackage{multirow}

\makeatletter
\renewcommand*{\fnum@figure}{{\normalfont \small{FIG.}~\thefigure}}
\makeatother

\makeatletter
\def\@seccntformat#1{\csname the#1\endcsname\quad}
\renewcommand\thesection{\arabic{section}}
\renewcommand\thesubsection{\thesection.\Alph{subsection}}
\renewcommand\thesubsubsection{\thesubsection.\arabic{subsubsection}}
\makeatother

\begin{document}

\title{Multimodal motion and behavior switching of multistable ciliary walkers}

\author{Sumit Mohanty}
\altaffiliation{These authors contributed equally to this work}

\affiliation{Autonomous Matter Department, AMOLF, Science Park 104, Amsterdam, 1098 XG The Netherlands}
\affiliation{Institute for Complex Molecular Systems and Department of Mechanical Engineering, Eindhoven University of Technology, P.O. Box 513, Eindhoven, 5600 MB The Netherlands}

\author{Paul Baconnier}
\altaffiliation{These authors contributed equally to this work}

\affiliation{Autonomous Matter Department, AMOLF, Science Park 104, Amsterdam, 1098 XG The Netherlands}

\author{Harmannus A. H. Schomaker}
\affiliation{Autonomous Matter Department, AMOLF, Science Park 104, Amsterdam, 1098 XG The Netherlands}

\author{Alberto Comoretto}
\affiliation{Autonomous Matter Department, AMOLF, Science Park 104, Amsterdam, 1098 XG The Netherlands}

\author{Martin van Hecke}
\affiliation{Autonomous Matter Department, AMOLF, Science Park 104, Amsterdam, 1098 XG The Netherlands}
\affiliation{Huygens-Kamerlingh Onnes Laboratory, Leiden University, 2300 RA Leiden, The Netherlands}

\author{Johannes T.B. Overvelde}
\affiliation{Autonomous Matter Department, AMOLF, Science Park 104, Amsterdam, 1098 XG The Netherlands}
\affiliation{Institute for Complex Molecular Systems and Department of Mechanical Engineering, Eindhoven University of Technology, P.O. Box 513, Eindhoven, 5600 MB The Netherlands}

\begin{abstract}
The collective motion of arrays of cilia - tiny, hairlike protrusions - drives the locomotion of numerous microorganisms, enabling multimodal motion and autonomous switching between gaits to navigate complex environments. To endow minimalist centimeter-scale robots with similarly rich dynamics, we introduce millimeter-scale flexible cilia that buckle under the robots weight, coupling multistability and actuation within a single physical mechanism. When placed on a vibrating surface, these ciliary walkers select their propulsion direction through the buckled states of their cilia, allowing multimodal motion and switching between modes in response to perturbations. We first show that bimodal walkers with left-right symmetric cilia can autonomously reverse direction upon encountering obstacles. Next, we demonstrate that walkers with isotropic cilia exhibit both translational and rotational motion and switch between them in response to environmental interactions. At increasing densities, swarms of such walkers collectively transition from predominantly spinning to translational motion. Finally, we show that the shape, placement and number of cilia controls the modes of motion of the walkers. Our results establish a rational, physically grounded strategy for designing minimalist soft robots where complex behaviors emerge from feedback between internal mechanical states and environmental interactions, laying the foundation for autonomous robotic collectives without the need for centralized control.
\end{abstract}

\maketitle

\section*{Introduction}

A key challenge in robotics is to emulate the rich locomotion of living organisms in non-living systems \cite{bechinger2016active, aguilar2016review, calisti2017fundamentals,po2024cooperative}. While advanced sensing and computing can endow robots with diverse gaits \cite{poulakakis2005modeling,raibert2008bigdog,spenko2008biologically,wooden2010autonomous}, it is striking that microorganisms such as \emph{placozoans} \cite{bull2021ciliary}, \emph{euplotes} \cite{Laeverenz-Schlogelhofer2024BioelectricEuplotes}, and bacteria \cite{sourjik2012responding, Vater2014SwimmingTracking, Hintsche2017ABody} naturally exhibit multiple gaits and autonomously switch between them in response to environmental cues - all without a central nervous system. Inspired by these behaviors, multimodal motion has recently been achieved by engineering soft robots, that use, e.g.,
coupled fluidic oscillators \cite{Comoretto2025PhysicalLocomotion, comoretto2025embodying} or mechanical linkages between legs and compliant antennas \cite{kamp2024reprogrammable} to control and switch between distinct gaits. Alternatively, multimodal motion can also emerge in, e.g., vibrated droplets and bubbles \cite{couder2005walking, valani2019superwalking, Guan2025GallopingBubbles}, or in flexible robots whose shape adapts to external cue to allow navigation through a maze \cite{xi2024emergent}. These examples show that multimodal motion can arise from only a few ingredients, but a unifying framework is lacking and it is difficult to design and control the modes of propulsion. Can a general, physics-based strategy enable minimalist robots to switch between modes, without resorting to sensors, actuators and centralized computing?

To address this challenge, we integrate actuation, feedback, and multistability through {\em flexible cilia}, extending cilia-based propulsion with buckling dynamics. Dense arrays of cilia power microorganisms such as \emph{Trichoplax Adhaerens}, where coordinated cilia beating drives multiple modes of rotational and translational locomotion \cite{bull2021ciliary}, and whose responses to environmental cues arise from passive mechanical compliance \cite{Smith2019CoherentSystem, gilpin2020multiscale, Davidescu2023GrowthSystem}.
We build upon centimeter-scale bristle bots, which rely on relatively stiff cilia and are powered by vibrations, and widely used as self-propelled synthetic walkers \cite{giomi2013swarming, ben2023morphological, novkoski2025graspion}. However, in contrast with living organisms, they exhibit a single mode of forward motion, determined by the symmetry-broken slant of their cilia \cite{reis2016vibration, cicconofri2016inversion, majewski2017locomotion}. Inspired by these systems, our key idea is to consider flexible symmetric cilia that under weight adopt different collective buckled configurations, endowing the array with multistability and thereby embodying distinct internal states.

We show that when a walker with such flexible cilia is subjected to vertical vibration, each cilium generates a propulsive force determined by its buckled configuration. At the same time, the motion of the walker impacts the buckling directions, thus setting up a two-way feedback loop between motion and internal state, similar to that of active materials \cite{ferrante2013elasticity, boudet2021collections, baconnier2022selective, baconnier2023discontinuous, Son2025EmergentLink-bots}. In our system, this feedback gives rise to multiple distinct modes of motion, which reflect the collective buckling modes of the cilia array.
Crucially, mechanical interactions with the environment - such as collisions with boundaries - can momentarily perturb that motion and reorient the buckled cilia, enabling transitions between the modes of propulsion, and thus allowing our walkers to autonomously change behavior upon reaching a boundary. This strategy integrates feedback between motion, multistability and environmental interaction within the cilia, providing a minimalist route to autonomous robots based on emergent, adaptive motion.

\begin{figure*}[t!]
\centering
\includegraphics[width=17.8cm]{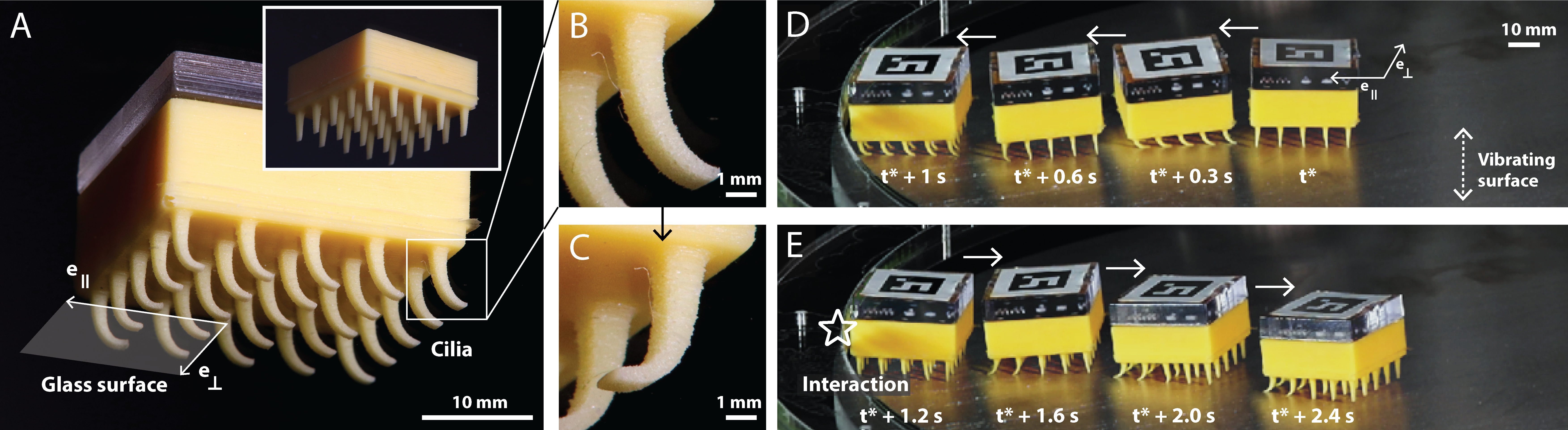}
\vspace{-0.2cm}
\caption{\textbf{Bimodal motion of ciliary walkers with rectangular-section cilia}. (A) A ciliary walker supported by an array of flexible cilia (inset); view from below a transparent glass plate. Each cilium is straight when unloaded (inset), and has a rectangular cross-section, enabling two distinct stable buckled states, i.e. the left- (B) and right-leaning (C) configurations (along and opposite to $e_{\parallel}$). (D) Superimposed snapshots of a walker translating on the vibrated plate, and (E) autonomously reversing direction upon collision with the boundary, exploiting the bistability of the buckled cilia.}\label{fig:1}
\end{figure*}

Combining experiments and simulations, we systematically investigate the motion of vertically-vibrated ciliary walkers. We start from bimodal walkers based on left-right buckling cilia, and demonstrate that these can reverse motion when reaching a boundary. Moreover, we show that this behavior can be understood at the level of a single cilium, as well as in mesoscopic models of collections of cilia with appropriate feedback between state and motion. We then introduce multimodal walkers, that are based on isotropic cilia. We show that these can switch between translational and spinning motion, and we rationalize these from the collective buckling-motion feedback in a mesoscopic model. We show that the location, shape and orientation of the cilia unlock a wide repertoire of rationally designed multimodal motions. Finally, we explore the collective behavior of swarms of walkers in crowded environments and reveal how their density affects their motion. Our work provides a scalable, physically grounded approach to designing robotic propulsion that emulates - and builds upon - the capabilities of natural locomotors. Altogether, it demonstrates the power of soft robotic systems in which complex autonomous behaviors emerge from interactions between internal degrees of freedom and the environment.

\section*{Bimodal motion of bidirectional walkers}

We first aim to realize bidirectional walkers that move either left or right. A natural strategy to do so is to employ bistable cilia, which, at rest, buckle left and right. Under vertical vibration, these generate a corresponding propulsive force, thereby enabling left and right motion.

Our walkers feature a cuboidal upper body supported by an array of millimeter-scale cilia which collectively buckle under the weight of the walker (Fig.~\ref{fig:1}A, \emph{SI Appendix}, section 1.A). We use tapered cilia featuring {\em anisotropic} rectangular sections that are aligned with
the local reference frame $(e_{\parallel},e_{\perp})$
of the walker, thereby restricting buckling to the 
$\pm e_{\parallel}$ directions (Figs.~\ref{fig:1}B-C). Finally, we place the cilia in a $5 \times 5$ square array.

To study their locomotion, we place our walkers on a vertically vibrated plate that provides driving and control the dimensionless peak acceleration $\Gamma = A (2 \pi f)^2 / g$, where $A$ is the vibration amplitude, $f$ its frequency, and $g$ gravity (\emph{SI Appendix}, section 1.B) \cite{cicconofri2016inversion, lanoiselee2018statistical, briand2018spontaneously}. For sufficiently strong vibrations, the bidirectional walkers spontaneously move along the direction of their buckled cilia, i.e., along $+e_{\parallel}$ or $-e_{\parallel}$ (SI Movie 1). Strikingly, the directions of motion and buckling can jointly reverse, for instance after colliding with the arena boundaries (Figs.~\ref{fig:1}D-E). Hence, the bilateral buckling of the cilia allows to realize walkers that can switch between two directions of motion.

\begin{figure*}
\centering
\includegraphics[width=17.8cm]{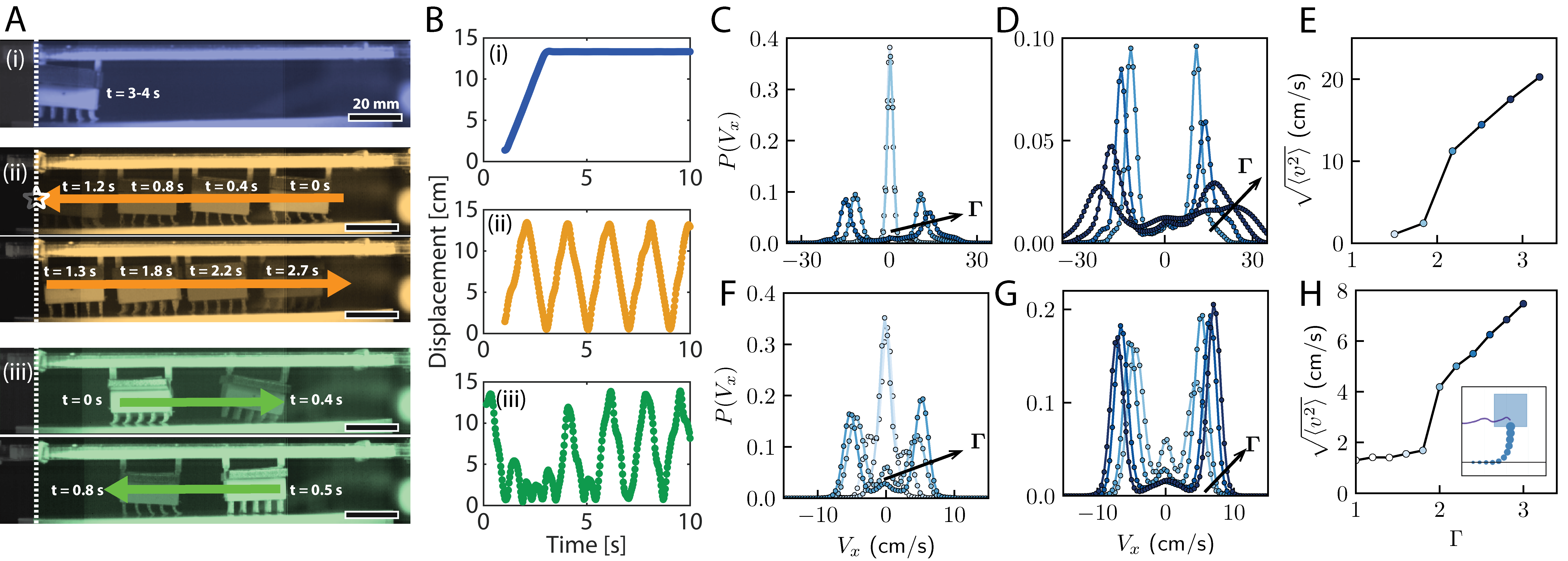}
\vspace{-0.25cm}
\caption{\textbf{Different types of locomotion in a 1D track.} (A) Time-lapse sequences (side view) of a bidirectional walker moving along a 1D track under three distinct regimes. (i) For $\Gamma = 1.97$, $W=12$ g, the walker exhibits steady motion until it reaches a boundary and then gets \textit{trapped}. (ii) For $\Gamma = 2.45$, $W=12$ g, the walker \textit{reverses} its motion upon collision with a boundary. (iii) For $\Gamma = 2.93$, $W=8$ g, the motion becomes \textit{erratic}, allowing spontaneous reversal of motion. 
(B) Corresponding trajectories. (C) Probability distributions of the velocity $V_x$, color coded from light to dark blue as the vibration amplitude $\Gamma \in \{ 1.49, 1.97, 2.45 \}$
increases illustrating the transition from the \textit{trapped} to the \textit{reversing} regime.
(D) When $\Gamma$ is increased from $1.97$ to $3.41$, the cruise velocity $v_0$ increases, and for strong driving the motion becomes erratic. (F-G) Corresponding PDFs for simulations of a single cilium (see inset of H and Supporting Information). (E and H) Second moment of the velocity $V_x$ as a function of $\Gamma$ for experiments (E) and for simulations of a single cilium (H).
}
\label{fig:2}
\end{figure*}

\subsection*{Locomotion behavior of bidirectional walkers}

The bimodal motion of bidirectional walkers gives rise to distinct types of locomotion behaviors, distinguished by their interactions with obstacles. We investigate these behaviors by confining the walkers in a $15$ cm long one-dimensional track and measuring the velocity $V_x$ along the track, keeping $f = 40$ Hz fixed and varying $\Gamma$ (Figs.~\ref{fig:2}A).
For weak vibrations ($\Gamma \lesssim 1$), the walker remains {\em static}, while for stronger vibrations, we observe three different locomotion behaviors. {\em(i)} When $1 \lesssim \Gamma \lesssim 2$, the walker moves left or right with a constant cruise speed $v_0=|\langle V_x \rangle|$, and its direction follows the configuration of the cilia. 
High-speed imaging reveals that this behavior originates from stick-slip motion of the cilia, synchronized with the driving (\emph{SI Appendix}, Fig.~S4.A-iii, SI Movie 3). When the walker encounters the edges of the track it simply stops moving and gets \textit{trapped} (Figs.~\ref{fig:2}A-i and \ref{fig:2}B-i, SI Movie 2). (ii) For larger vibration amplitudes $2 \lesssim \Gamma \lesssim 3$, the walker moves left or right and reverses its direction of motion upon collision with the boundaries (Fig.~\ref{fig:2}A-ii, SI Movie 2). This \emph{reversing} locomotion results in back and forth motion between the two ends of the track (Fig.~\ref{fig:2}B-ii). 
(iii) When $\Gamma$ is increased further, the motion becomes increasingly {\em erratic}, and exhibits spontaneous reversal where the walker switches between left and right buckled states (Figs.~\ref{fig:2}A-iii and \ref{fig:2}B-iii, SI Movie 2). In addition, the motion becomes so energetic that the walker may escape the track.

To further investigate the \textit{trapped} and \textit{reversing} regimes, we track the motion of the walker, determine its velocity $V_x(t)$, and study its probability distribution function (PDF) as function of $\Gamma$.
First, we observe a nearly sharp transition from the trapped regime, where the PDF is sharply peaked around $V_x=0$, to the reversing regime where the PDF is bimodal, reflecting the back and forth motion of the walker (Fig.~\ref{fig:2}C). Second, in the reversing regime, we observe that the speed $v_0$ increases with $\Gamma$, and that the peaks broaden and become asymmetric, reflecting the smooth crossover to the erratic regime (Fig.~\ref{fig:2}D). Plotting the second moment $\langle V_x^2\rangle$ as function of $\Gamma$ confirms the sharp transition between the trapped and reversing regime, and the smooth increase of the velocity for large $\Gamma$ (Fig.~\ref{fig:2}E). 

Together, this shows that the bimodal motion of these walkers, driven by collective left-right buckling of their cilia, produces distinct locomotion regimes, including autonomous reversals that allow to avoid obstacles and boundaries. The bistability of the cilia, coupled with feedback between cilia motion and buckling direction, enables passive walkers to exhibit seemingly intelligent locomotion.

\subsection*{Single cilium modeling}
 
We now investigate whether a single cilium exhibits similar bimodal motion and locomotion regimes. We study a simplified model of a vibrated buckled cilium, consisting of a discrete chain of linear and torsional springs calibrated to reproduce the force-displacement curve of an experimental cilium under compression tests (\emph{SI Appendix}, Fig.~S3.A-B). The chain's top carries a load and is constrained to be vertical, while its bottom rests on the vibrating plate and interacts frictionally with it as it moves along the track (Fig.~\ref{fig:2}H-inset). Strikingly, this model reveals the same sequence of regimes as in experiments (\emph{SI Appendix}, section 2). For $\Gamma < 1$ the buckled cilium remains static. Then, at larger $\Gamma$, the buckled cilium moves left or right depending on its buckled configuration. For low $\Gamma$ it gets trapped by the boundaries, thus the velocity distribution peaks at $V_x = 0$ (Fig.~\ref{fig:2}F, \emph{SI Appendix}, Fig.~S5B, SI Movie 4). At larger $\Gamma$, the cilium gains sufficient momentum to land upright after the collisions, allowing a transition to the opposite buckled state and reversal of motion, which leads to a bimodal velocity distribution (Figs.~\ref{fig:2}F-G, \emph{SI Appendix}, Fig.~S5A, SI Movie 4). Even larger $\Gamma$ lead to a smooth increase of the velocity (Fig.~\ref{fig:2}H), while the motion becomes progressively more erratic. 

Hence, this model suggests that reversal can be understood from the straightening of the cilia during collisions, and that the bidirectional motion and locomotion regimes observed for walkers with many cilia follow from the dynamics of a single cilium.

\begin{figure*}[t!]
\centering
\includegraphics[width=17.8cm]{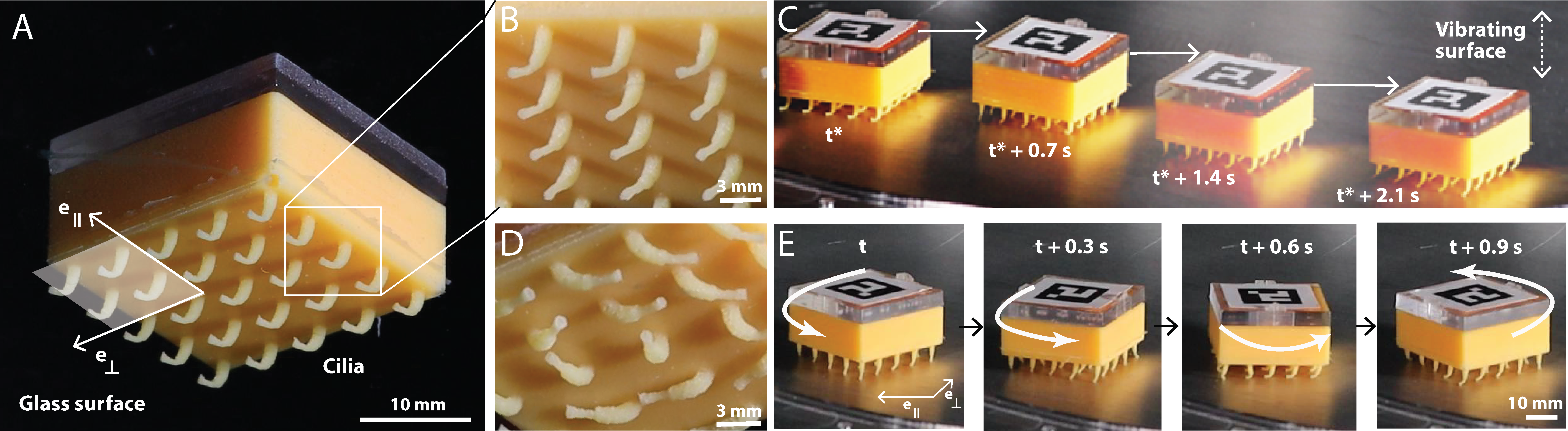}
\vspace{-0.2cm}
\caption{\textbf{Multimodal motion of isotropic ciliary walkers.} (A) A ciliary walker supported by an array of flexible cilia; view from below a transparent glass plate. Each cilium has a cylindrical cross-section, enabling multiple distinct buckled states at the level of the cilia array, i.e. translational (B) and rotational (D) configurations. (C-E) Superimposed snapshots of a walker performing translational (C) and rotational (E) motion on the vibrated plate.}
\label{fig:3}
\end{figure*}

\subsection*{Collective organization of bistable cilia} 

The left or right buckled state of the cilia determines their motion, while in turn, motion of the walker (by self-propulsion, external manipulation, or collisions with the boundary) can flip the state of the cilia and reverse the direction of propagation. To understand this feedback between motion and internal state at the collective level, we performed additional simulations on a distinct model of bidirectional walkers, inspired by recent works on elastically-coupled active units \cite{baconnier2022selective, hernandez2024model}. These are composed of an array of rigidly coupled active units which can be left or right polarized and which represent the cilia. To model the feedback, each unit exerts a force on the supporting plane aligned with its buckling direction, while its left-right polarization reorients toward the collective motion of the walker through a generic mechanism called \textit{self-alignment} \cite{baconnier2025self} (\emph{SI Appendix}, section 3.C).
We find that, starting from an initial state with random polarizations, the system quickly organizes into a fully polarized state with velocity $V_x = \pm v_0$, and that external forces are able to change this state, reversing the direction of propagation (\emph{SI Appendix}, Figs.~S7 and S8). Moreover, we demonstrate that, in the case of purely elastic collisions at the boundaries, the autonomous reversals upon collisions only arise in the presence of inertia (\emph{SI Appendix}, Figs.~S8H). 

Altogether, the feedback between motion and state, already present at the level of a single cilium, persists in collections of cilia that reorient toward their motion.

\section*{Multimodal motion of isotropic walkers}

We now show that the shape of the cilia can be leveraged to control the nature and multiplicity of the modes of motion of the walkers. Specifically, we now consider cilia with circular cross-sections which can buckle in any direction in the plane, and place these in a $5 \times 5$ square array to 
create \textit{isotropic walkers} (Fig.~\ref{fig:3}A).

We find that these walkers can propagate with constant speed in any direction with respect to the orientation of their body, and that gentle nudges
can set the walker into motion in a different direction (Figs.~\ref{fig:3}B-C). As expected, the directions of buckling and propagation are following each other during translational motion. Strikingly, by imposing a twist to the walker, 
the cilia can buckle into a vortex-like pattern, resulting in clockwise or counter-clockwise spinning motion (Figs.~\ref{fig:3}D-E, SI Movie 5). 
Because the cilia array can buckle collectively in any direction or adopt chiral configurations, two distinct groups of motion modes emerge, and the 
propagation modes remain closely linked to the collective buckling modes.

\begin{figure*}[t!]
\centering
\includegraphics[width=17.8cm]{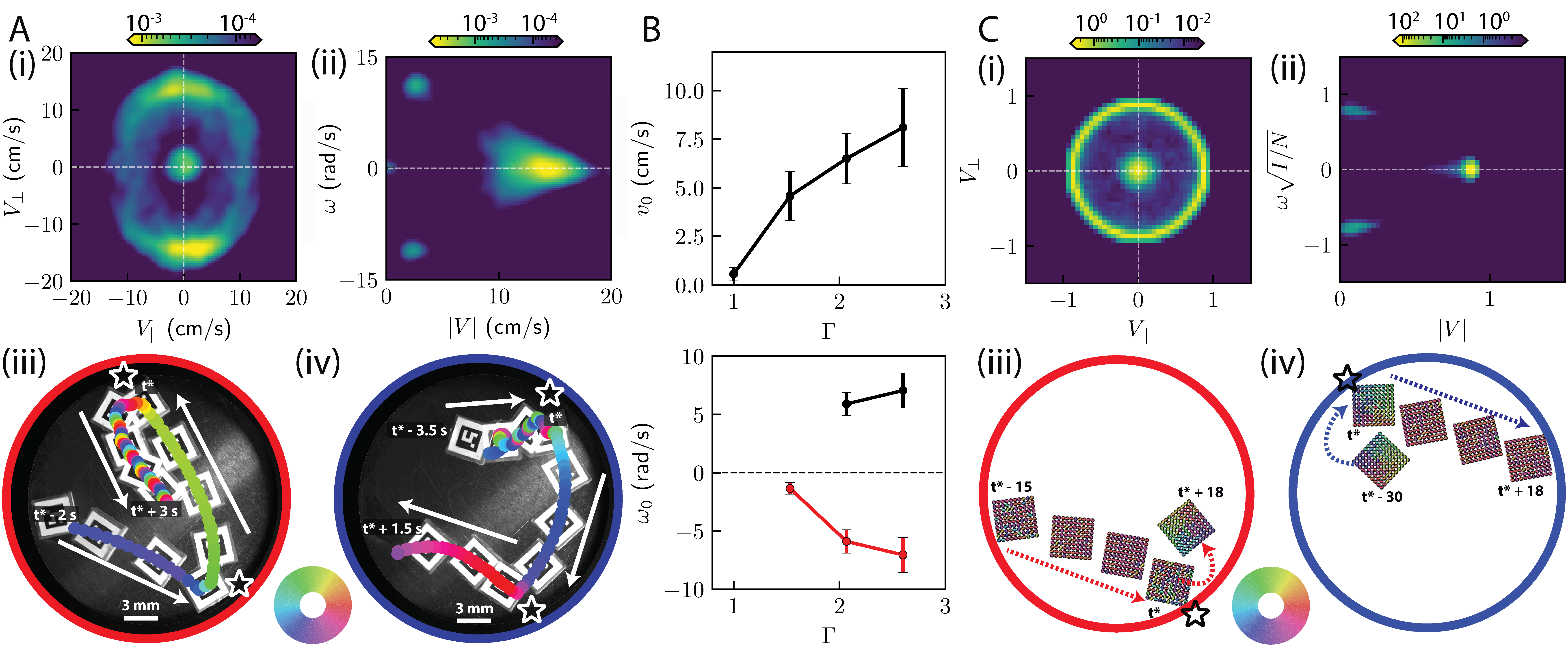}
\vspace{-0.1cm}
\caption{\textbf{Transitions between locomotion modes for isotropic ciliary walkers in a 2D arena.} (A) Probability distributions of the rotation rate and velocity, as obtained from experiments at fixed $f = 60$ Hz, $\Gamma = 5.05$, $W = 12$ g; in the $V_{\parallel}-V_{\perp}$ plane (i), and in the $|\boldsymbol{V}|-\omega$ plane (ii), overlaid with representative trajectories from (iii-iv); arrows indicate the direction of the transitions between locomotion modes. (iii-iv) Superimposed snapshots of a walker during collisions with a boundary, showing autonomous transitions from translational to rotational (iii) and from rotational to translational motion (iv); the colored markers indicate the orientation of the velocity $\boldsymbol{V}$ of the walker (inset colorbar), and collisions are marked with a star symbol. (B) Coordinates of the detected peaks from the experimental probability distributions as a function of $\Gamma$, as obtained from experiments at fixed $f = 40$ Hz, $W = 12$ g. (C) Same as (A), as obtained from simulations of a rigid active solid, with fixed $D = 0.2$, $\tau_n = 0.5$, $M = 1$. In the bottom snapshots, the arrows represent the active forces and are color coded according to their orientation (inset colorbar).
}
\label{fig:4}
\end{figure*}

\subsection*{Locomotion in a 2D arena}

We now examine the statistical properties of the motion of an isotropic walker in a disk-shaped arena ($16$ cm in diameter), focusing on how interactions with boundaries lead to transitions between translational and spinning modes of motion. We characterize the motion by the probability distributions as a function of
$V_{\parallel}, V_{\perp}$ and of the rotation rate $\omega$. While for small enough $\Gamma$, we again observe a regime where the walkers get trapped, we focus on larger $\Gamma$ values where they bounce back from the boundaries. 

We first show that, in contrast to bidirectional walkers, the translational motion is mostly decoupled from the orientation of the walker: the PDF projected on the ($V_{\parallel}, V_{\perp}$) plane is nearly isotropic (Fig.~\ref{fig:4}A-i). We do, however, observe that walkers tend to favor motion parallel to the $e_{\perp}$ direction. We attribute this to fabrication anisotropy in the 3D-printed molds, which produce slight anisotropies in the cilia cross-sections and concomitant deviations from isotropy in buckling directions (\emph{SI Appendix}, section 1.A).

We then investigate the competition between translation and spinning from the PDF projected on the $|\boldsymbol{V}|-\omega$ plane. We can clearly distinguish distinct motion modes (Fig.~\ref{fig:4}A-ii). First, the peak at zero spin and finite speed corresponds to the continuous spectrum of translational modes discussed above. Second, we find two speaks at positive and negative spin, and small velocity. These correspond to two discrete spinning modes, where we attribute the small translational component to finite number effects: e.g., in the $5\times5$ array, the propulsive force exerted by the central cilium breaks the polar symmetry of the structure, thus preventing pure rotational motion. 
When propagating in the 2D arena, the walkers frequently collide with the boundaries. Such collisions can lead to transitions between translational and spinning propagation modes (Fig.~\ref{fig:4}A-iii,iv, \emph{SI Appendix}, Fig.~S6, SI Movie 6).

Next, we characterize the effect of the vibration amplitude on the properties of the motion modes. We thus measure the cruise velocity $v_0$ and rotation rates $\omega_0^{\pm}$ as a function of $\Gamma$, where $v_0$ and $\omega_0^{\pm}$ are obtained from the maxima of the PDFs of $|\boldsymbol{V}|$ and $|\omega|$, respectively (\emph{SI Appendix}, section 1.D.3). First, we find that $\omega_0^{-}$ and $\omega_0^{+}$ are nearly equal and opposite (Fig. \ref{fig:4}B). Second, both the cruise velocity of the translational modes $v_0$ and the typical rotation rates of the rotational modes $\omega_0^{\pm}$ exhibit a transition from zero to finite values, and then increase with $\Gamma$ (Figs. \ref{fig:4}B). We also note that these onsets occur for different values of $\Gamma$, leaving a small interval where the motion is almost purely translational, and that a detailed analysis of the amplitudes of the peaks shows that, for this design, increasing $\Gamma$ favors translational motion over rotational motion.

\subsection*{Collective organization of isotropic cilia}

The feedback mechanism between the motion and the buckled states of isotropic cilia is strongly reminiscent of the physics of elastically-coupled active units that have been studied intensely recently
\cite{baconnier2022selective, hernandez2024model}. Here we simulate such a model and show that it captures the experimentally observed behavior of isotropic walkers in confinement (\emph{SI Appendix}, section 3, SI Movie 7)

We consider isotropic walkers as arrays of rigidly coupled but freely-rotating active units which 
represent the buckled cilia, and which exert polar forces in the plane along the direction of the polarity $\boldsymbol{\hat{n}}_i$ \cite{baconnier2022selective, hernandez2024model}. Individually, the active units move with a velocity $\boldsymbol{v}_i = V_0 \boldsymbol{\hat{n}}_i$, where $V_0$ results from the balance between self-propulsion and friction. Moreover, the polarity $\boldsymbol{\hat{n}}_i$ reorients toward the velocity of the unit $\boldsymbol{v}_i$ according to a self-alignment mechanism \cite{baconnier2025self}. 
Choosing the units of time and length such that $r_0 = l_0$ and $t_0 = l_0/V_0$, with $l_0$ the inter-cilia distance, the dimensionless equations of motion can be written as \cite{baconnier2022selective, hernandez2024model}:
\begin{subequations} \label{eq:rigid_active_solids_general}
\begin{align}
 M \frac{d \boldsymbol{v}_i }{dt} &= - \boldsymbol{v}_i + \boldsymbol{\hat{n}_i} + \boldsymbol{F}_i^{el}, \label{eq1:rigid_active_solids_general} \\
 \frac{d \boldsymbol{\hat{n}}_i }{dt} &= \frac{1}{\tau_n} (\boldsymbol{\hat{n}}_i \times \boldsymbol{v}_i) \times \boldsymbol{\hat{n}}_i + \sqrt{\frac{2D}{\tau_n}} \xi_i \boldsymbol{\hat{n}}_i^{\perp}, \label{eq2:rigid_active_solids_general} 
\end{align}
\end{subequations}
where $\boldsymbol{F}_i^{el}$ is the elastic force exerted by the rigid matrix on agent $i$ (\emph{SI Appendix}, section 3). The model contains three dimensionless parameters: (i) the alignment $\tau_n = l_a/l_0$, where $l_a$ is the alignment length over which $\boldsymbol{\hat{n}}_i$ aligns with $\boldsymbol{v}_i$; (ii) the amplitude of angular noise $D$; and (iii) the momentum $M = m V_0 / \gamma l_0$, where $m$ is the agent mass and $\gamma$ is an effective friction coefficient (\emph{SI Appendix}, section 3). Eq.~(\ref{eq2:rigid_active_solids_general}) governs the reorientation of the active units, and contains the nonlinear self-alignment term and a noise term with correlations $\langle \xi_i (t) \xi_j (t') \rangle = \delta_{ij} \delta (t - t')$, while Eq.~(\ref{eq1:rigid_active_solids_general}) governs the balance between active, elastic, and friction forces, and, in contrast to earlier work, includes inertial effects $\propto$ $M$ \cite{caprini2022role, fersula2024self}.
Without inertia ($M = 0$) and for small enough noise ($D < 0.5$), this model gives rise to emergent collective rotational and translational motions \cite{hernandez2024model}.

To mimick the motion of isotropic walkers, we simulate this model for large enough $M$ in a disk-shaped arena and include elastic collisions with the boundaries (\emph{SI Appendix}, section 3.D). As expected, the motion statistics reveals
an isotropic distribution of translational modes and two rotational modes with opposite handedness
(Fig. \ref{fig:4}C-i,ii).
Strikingly, collisions with the boundary can give rise to transitions between different translational modes, or between translational and rotational modes, as observed experimentally (Figs. \ref{fig:4}C-iii,iv, \emph{SI Appendix}, Fig. S9). Hence, while details of the motion statistics and switching behavior depend on the specifics of the experimental implementation, a simplified active solid model with purely elastic collisions
already reproduces all experimentally observed behaviors, suggesting that the
mechanisms of feedback between cilia polarizations and propulsive force, and the switching between modes of motion, is generic and robust.

\begin{figure}[t!]
\centering
\includegraphics[width=8.7cm]{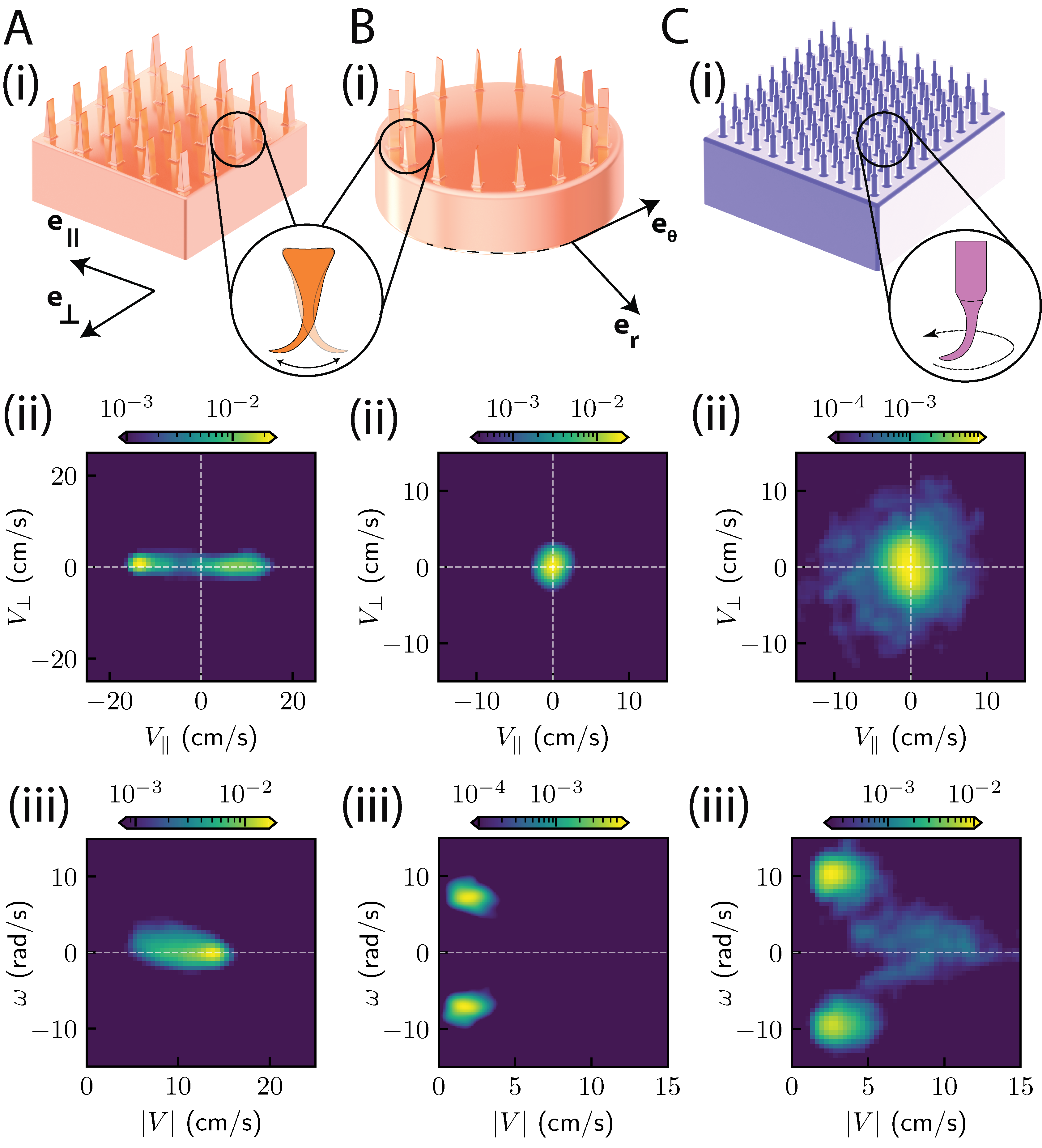}
\vspace{-0.65cm}
\caption{\textbf{Tunable locomotion behaviors in a 2D arena using different designs of cilia arrays.} (i) Schematics of the different arrays of flexible cilia; A: $5 \times 5$ square array with bistable, rectangular-section cilia, all oriented along $\boldsymbol{\hat{e}}_{\parallel}$; B: circular line of bistable, rectangular-section cilia, all oriented along $\boldsymbol{\hat{e}}_{\theta}$; C: $10 \times 10$ square array with multistable, cylindrical-section cilia. (ii) Different arrays of cilia give rise to drastically different locomotion behaviors, as illustrated in the planes $V_{\parallel}-V_{\perp}$ and $|\boldsymbol{V}|-\omega$ (same conventions as Fig. \ref{fig:4}A-i,ii); as obtained from experiments at fixed $f = 40$ Hz, and $\Gamma = 2.45$,$2.45$, and $1.49$, respectively. Added weight on walkers in A and C was $8$ g.}
\label{fig:5}
\end{figure}

\subsection*{Programming the modes of motion}

We now show how the shape, placement and number of cilia can be used to program the modes of motion of the walkers in experiments.

First, we revisit the bidirectional walker, but now let it freely move in the disk-shaped arena (Fig.~\ref{fig:5}A-i).
As the PDFs show, the motion is predominantly translational along and opposite to $\boldsymbol{\hat{e}}_{\parallel}$, even without a guide-rail, and rotational motion is suppressed (Figs.~\ref{fig:5}A-ii,iii, SI Movie 8). For increasing vibration strength, the cruise velocity $v_{0}$ increases while the rotation rate remains small (\emph{SI Appendix}, Fig.~S11A).
Second, we place cilia with rectangular-cross sections in a circle and orient them azimuthally (i.e., along $e_{\theta}$, see Fig.~\ref{fig:5}B-i). This results in clockwise and counter-clockwise spinning motion, with a small translational component that we attribute to a slight tilt of the vibrated plate with respect to the horizontal (Figs.~\ref{fig:5}B-ii,iii, SI Movie 8). Moreover, the cruise rotation rates $\omega_0^{\pm}$ also increase with larger $\Gamma$ (\emph{SI Appendix}, Fig.~S11B). These two examples demonstrate that the shape and placement of the cilia can be used to select locomotion modes along different rigid body motions.

Finally, we investigate the role of the number of cilia $N$, by creating a $10 \times 10$ isotropic walker (Fig.~\ref{fig:5}C-i). This produces a walker that remains in an unbuckled configuration at rest, but where the cilia buckle under oscillations, leading to a broader distribution of translational velocities, alongside two dominant spinning modes (Figs. \ref{fig:5}C-ii,iii, SI Movie 8). Increasing $\Gamma$ enhances both the rotational rate $\omega_0$ and cruise velocity $v_0$ (\emph{SI Appendix}, Fig.~S11C), and shifts the overall probability distribution towards translational modes. Altogether, these results provide a glimpse of the large repertoire of locomotion behaviors of ciliary walkers that can be explored by varying the design and spatial placement of the cilia and controlling the vibration strength.

\section*{Collective behavior of multiple walkers}

\begin{figure}[t!]
\centering
\includegraphics[width=8.7cm]{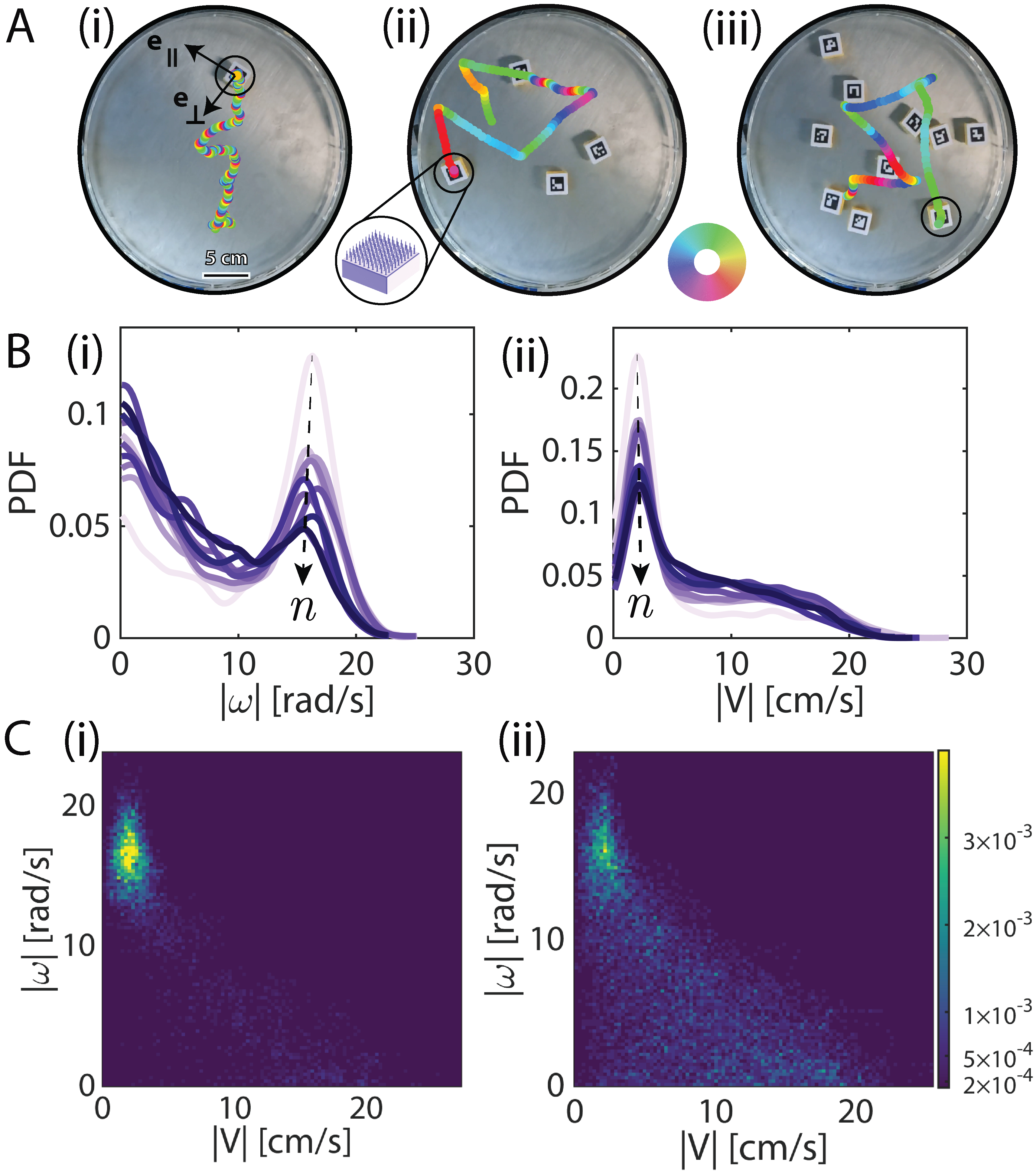}
\vspace{-0.6cm}
\caption{\textbf{Density-dependent locomotion behavior of multiple ciliary walkers.} (A) Multiple walkers with cylindrical-section cilia ($10 \times 10$ square array as in Figs. \ref{fig:5}C) interact through collisions in a disk-shaped arena, and constantly switch behaviors between translational and rotational modes (SI Movie 7); from left (i) to right (iii): snapshots for $n = 1, 4$, and $10$ walkers; the colored markers indicate the orientation of the velocity of a given walker (black circles); the overlaid trajectories correspond to the last (i) $12$ and (ii,iii) $5$ s of motion, respectively. (B) Probability distributions of the absolute value of the (i) rotational and (ii) translational components of a given walker for different number of walkers $n \in \left[ 1, \dots, 10\right]$, color-coded from light to dark blue as $n$ increases. (C) Motion statistics for (i) $n=1$ and (ii) $n=10$ in the plane $|V|-|\omega|$, where (i) and (ii) are associated to the lightest ($n=1$) and darkest ($n=10$) blue curves of B, respectively. In all panels, $\Gamma \simeq 1.5$ and $W = 0$ are fixed.}
\label{fig:6}
\end{figure}

We finally explore the interactions and collective behaviors of ciliary walkers in a swarm. We consider isotropic walkers based on $10 \times 10$ square arrays of isotropic cilia in a large disk-shaped arena ($36$ cm in diameter) at fixed $\Gamma \sim 1.5$ (Fig.~\ref{fig:6}A, \emph{SI Appendix}, section 1.B). 
For these parameters, a single walker exhibits both translational and rotational modes of motion. Here, the large arena gives rise to long-lasting spinning motion with only a small translational drift, which in turn dominates the PDF (Figs.~\ref{fig:6}B,C-i). Adding more walkers increases the frequency of collisions between walkers. As a result, the spinning mode of motion is less frequent, and the contribution of translational motion increases (Fig.~\ref{fig:6}B,C-ii, SI Movie 9). This trend can be understood by realizing that a collision between a spinning walker - with boundaries or other walkers - often leads to a transient translational motion. 
The walkers then recover their steady rotation rate $\pm \omega_0$ and low translational velocity
after a typical transient time $\tau$ (in the absence of subsequent collisions). We suggest that $\tau$
is governed by the collective dynamics of the cilia array. When the density increases, the walkers spend more and more of their time in this transient phase, so that at large enough density, the motion is dominated by translational motion.
This strong dependence of the individual motion statistics on the density of ciliary walkers within a swarm highlights how interactions among walkers enable both autonomous switching between
modes and density-dependent types of behaviors.

\section*{Discussion}

In this manuscript, we have introduced a class of vibrated ciliary structures that exhibit multiple locomotion modes and autonomously switch between them through physical interactions with their environments. Compared to previous work, e.g., networks of elastically-connected bristle bots \cite{baconnier2022selective, hernandez2024model, xi2024emergent}, our approach integrates the emergent functionality of multiple active units to a single monolithic walker.
Moreover, this versatility does not require sophisticated multi-component architectures, like on-board electronics or sensors, and thus should be relatively straightforward to translate to robotic approaches at sub-centimeter scales \cite{miskin2020electronically, ng2021locomotion}. Altogether, ciliary walkers demonstrate a scalable, physically grounded strategy to replicate the adaptive capabilities of biological locomotors \cite{bull2021ciliary, Peerlinck2023ArtificialNature}.

Beyond the specific design of ciliary walkers, this work opens avenues for exploring emergent autonomy in living and robotic systems, for example in other active systems with multiple coexisting dynamical states \cite{souslov2017topological, morin2018flowing, baconnier2023discontinuous, jorge2024active, jorge2025active}. Furthermore, ciliary walkers close the gap between robotics and active matter. From a robotics viewpoint where autonomy has traditionally been engineered through external control and centralized processes, here we place more emphasis on the power of harnessing emergent collective behaviors as a result of interactions between buckled cilia. From an active matter point of view, we demonstrate that harnessing buckling can lead to more complex behaviors at a single particle level.
Extending these ideas, ciliary arrays could be engineered to exhibit soft deformation modes \cite{baconnier2022selective, hernandez2024model} or frequency-selective locomotion \cite{Cui2023MiniaturizedCilia}, further broadening the functional space of our walkers. 

Our experiments with multiple ciliary walkers highlight the emergence of density-dependent behaviors, and reveal how inter-agent interactions alone can drive behavioral switching. It would be interesting to explore further if and how density-dependent behaviors can drive phase separation and self-organization phenomena at high density \cite{cates2015motility, ben2023morphological, Ziepke2025AcousticSystems}. These findings also suggest the potential for constructing soft robotics swarms capable of self-organizing into dynamic patterns and navigate complex environments, such as industrial sites or remote pipelines, driven solely by physical interactions, without external and centralized control. Looking ahead, the scalability of our platform opens a largely unexplored avenue in microrobotics, where such emergent autonomy may allow for operating in environments where traditional hardware-based control is infeasible. This untapped potential may inspire practical applications, from targeted clinical therapy \cite{Mohanty2020ContactlessSciences} to the development of active materials with living-like, animate properties \cite{Kriegman2020AOrganisms, veenstra2025adaptive}.

\section*{Acknowledgments}

JTBO acknowledges funding from the European Research Council Starting Grants no.$948132$). PB and MvH acknowledge funding from European Research Council Grant ERC no.$101019474$.  This work is part of the Dutch Research Council (NWO) and was performed at the research institute AMOLF. The authors thank Niels Commandeur and Dion Ursem in development of experimental setup, and Matthew S. Bull for sharing the insightful data on ciliary buckling of \emph{Trichoplax Adhaerens}.

\bibliography{refs.bib}

\appendix

\subsection*{Fabrication}

The ciliary walkers are fabricated using injection molding of prepolymer mixture of silicone elastomer (Elite Double 22F, Zhermack), with outer molds 3D-printed in VeroClear (Stratasys, Eden 260VS). The external weights to buckle the ciliary array are also made of VeroClear. More details on the fabrication can be found in \emph{SI appendix}, section 1.A.

\subsection*{Vibration platform}

All the experiments with the walkers are performed on a table-top vibrational platform comprising a flat aluminum plate mounted on an electrodynamic shaker (VTS systems VG100). In 1D experiments, the track boundaries consists of two stiff rubber bands at approximately the same height as the walkers body, while for the 2D arena, the boundary is a stiff material with a large circular hole at the same height as the walkers body. Calibration curves of the vibrated plate to determine the $\Gamma$ values, as well as additional experimental details, can be found in \emph{SI appendix}, section 1.B.

\subsection*{Tracking the motion of ciliary walkers}

Top-view recordings ($30$ fps) of ciliary walkers carrying printed ArUco markers (mounted on top weights) are analyzed using a custom Python routine based on \texttt{OpenCV}. ArUco markers are detected with \texttt{cv2.aruco.detectMarkers}, providing the centroid position $(x, y)$ and body orientation $\theta$. The instantaneous translational velocity $(V_x, V_y)$ and angular velocity $\omega = d\theta/dt$ are obtained from framewise displacements. The velocity components are smoothed using a Gaussian kernel and transformed into the local body frame defined by the orientation vectors obtained from ArUco markers $(\boldsymbol{e}_{\parallel}, \boldsymbol{e}_{\perp})$, yielding the parallel and perpendicular components $(V_{\parallel}, V_{\perp})$. For 1D locomotion, only the axial component $V_{\parallel}$ is retained for time-series and distribution analyses, while for 2D locomotion, the full set $(V_{\parallel}, V_{\perp}, \omega)$ is used to construct velocity histograms and perform subsequent statistical analyses.

\subsection*{Analysis of the probability distributions}

In 2D experiments, the cruise velocity $v_0$ and rotation rates $\omega_0^{\pm}$, if they exist, are determined by extracting the positions of the peaks in the distributions of $|\boldsymbol{V}|$ and $\omega$. We first represent the probability distribution of ($V_{\parallel}$, $V_{\perp}$,$\omega$) in the plane $|\boldsymbol{V}|-\omega$, where different lobes correspond to different locomotion modes. For isotropic walkers, the peak at finite $|\boldsymbol{V}|$ and vanishing $\omega$ corresponds to the translational modes, while the peaks at large $\omega$ and small $|\boldsymbol{V}|$ correspond to the two rotational modes. Then, we plot the probability distribution of $|V|$. In this case, we find two peaks: one at small velocity associated with the two rotational lobes, and one at larger velocity associated with the translational lobe. The latter maxima is used to define the cruise velocity $v_0$. Eventually, we plot the probability distribution of $\omega$, where we can clearly see the two rotational lobes with finite and almost opposite rotation rates. These two maxima are used to define $\omega_0^{\pm}$. The errorbars for these quantities represent the typical width of the peaks around $v_0$ and $\omega_0^{\pm}$, respectively. For a specific example, see \emph{SI appendix}, section 1.D.3.

\newpage
~\\
\newpage
\widetext
\begin{center}
\textbf{\large Supporting Information: Multimodal motion and behavior switching of multistable ciliary walkers} \\
\vspace{0.4cm}
Sumit Mohanty$^{\star,1,2}$, Paul Baconnier$^{\star,1}$, Harmannus A. H. Schomaker$^{1}$, Alberto Comoretto$^{1}$, Martin van Hecke$^{1,3}$, and Johannes T.B. Overvelde$^{1,2}$ \\
\vspace{0.2cm}
\textit{$^{1}$Autonomous Matter Department, AMOLF, Science Park 104, Amsterdam, 1098 XG The Netherlands.} \\
\textit{$^{2}$Institute for Complex Molecular Systems and Department of Mechanical Engineering, Eindhoven University of Technology, P.O. Box 513, Eindhoven, 5600 MB The Netherlands.}
\textit{$^{3}$Huygens-Kamerlingh Onnes Laboratory, Leiden University, 2300 RA Leiden, The Netherlands.}
\end{center}
\vspace{0.5cm}
\setcounter{equation}{0}
\setcounter{figure}{0}
\setcounter{table}{0}
\setcounter{page}{1}
\makeatletter
\renewcommand{\theequation}{S\arabic{equation}}
\renewcommand{\thefigure}{S\arabic{figure}}
\renewcommand{\thetable}{S\Roman{table}}
\renewcommand{\theHtable}{S\Roman{table}}   

\setcounter{section}{0} 
\renewcommand\thesection{\arabic{section}} 
\renewcommand{\theHsection}{\arabic{section}} 
\makeatletter
\renewcommand{\@seccntformat}[1]{\csname the#1\endcsname.\space}
\renewcommand{\appendixname}{} 
\def\p@section{} 
\def\p@subsection{} 
\def\thesection{\arabic{section}}
\renewcommand\thesubsection{\thesection.\Alph{subsection}} 
\renewcommand\thesubsubsection{\thesubsection.\arabic{subsubsection}}
\renewcommand{\@seccntformat}[1]{\csname the#1\endcsname.\space} 
\makeatother


\makeatother

\onecolumngrid

\noindent\textbf{This PDF file includes:} \\
- Supporting text \\
- Figs. S1 to S11 \\
- Table S11 \\
- Legends for Movies S1 to S9 \\
- SI References \\

\noindent\textbf{Other supporting materials for this manuscript include the following:} \\
- Movies S1 to S9 \\

\newpage

\section{Materials and Methods}

\subsection{Manufacturing} \label{sec:manufacturing}
\begin{figure}[p]
\centering
\includegraphics[width=\textwidth]{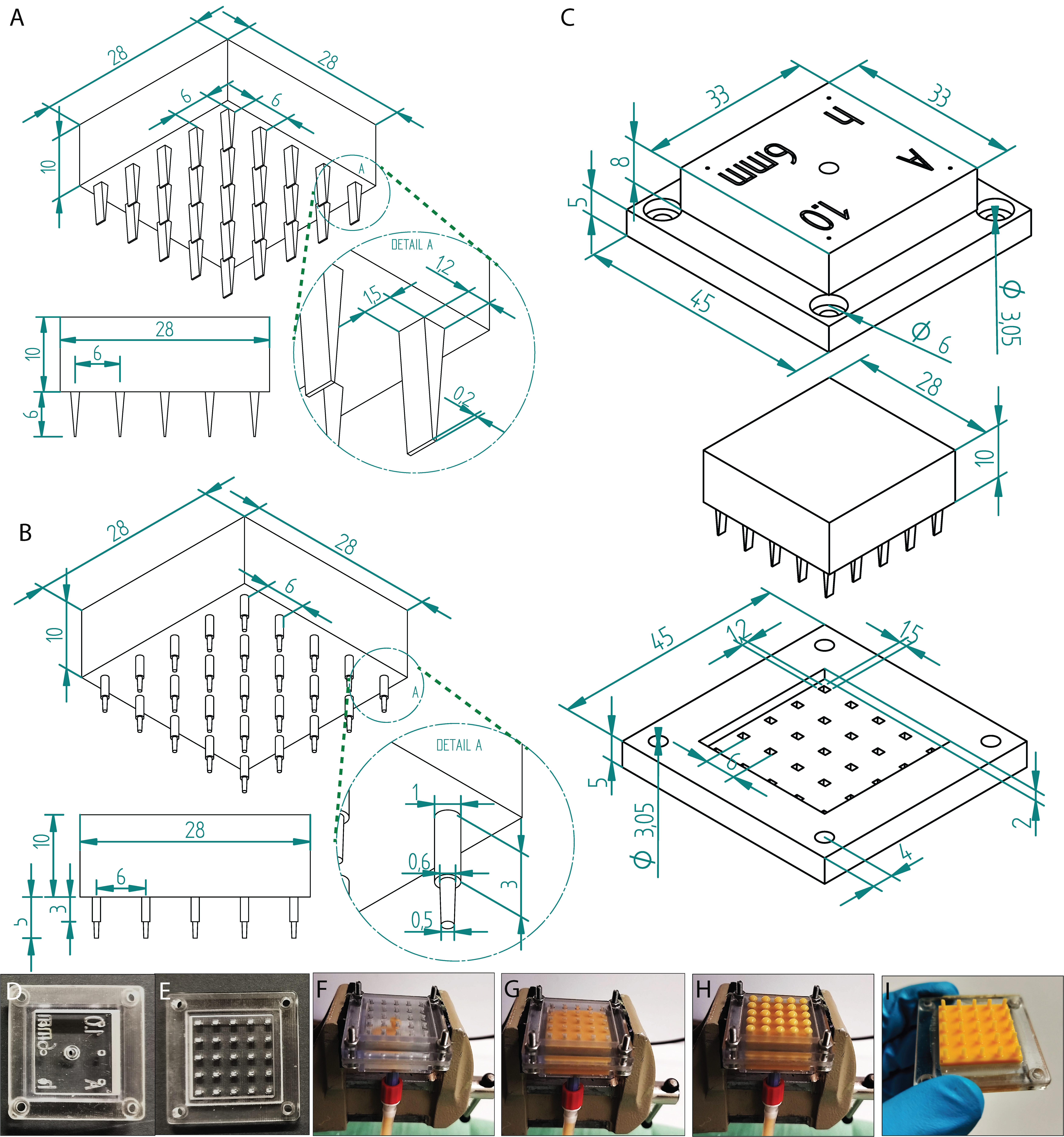}
\caption{Fabrication process of the ciliary walkers: drawings of the bidirectional (A) and isotropic (B) walkers with $5 \times 5$ square arrays. (C) Mold design and assembly for the bidirectional ciliary walkers. (D-E) Photographs of the mold post printing and development. (F-H) Injection molding procedure and (I) final sample highlighting the cilia array post curing and demolding process.} 
\label{fig:S1}
\end{figure}

The ciliary walkers and enlarged cilium are fabricated by injecting a fast-curing silicone elastomer (Elite Double, ED22F, Zhermack) into 3D-printed molds using a pneumatic dispenser gun. Each mold consists of two parts: (i) a top section with an inlet that forms a common cuboidal substrate ($27$ cm $\times$ $27$ cm $\times$ $10$ mm) with press-fit holes for external weights, and (ii) a bottom section with outlet channels that defines the geometry and distribution of the cilia, supporting different configurations (e.g., $N = 5 \times 5$ and $10 \times 10$). The two parts are screwed tightly together prior to injection, allowing silicone to fill the mold while air escaped through the outlets. This modular design enables the same substrate to be paired with interchangeable bottom molds for different walker variants. After curing, the molds are disassembled, and excess silicone on the inlet and outlet sides carefully trimmed with a blade. Detailed dimensions of the two types of walker, and the mold dimensions are provided in Fig. \ref{fig:S1}A-C. Note that isotropic walkers with $10 \times 10$ cilia use the same design as Fig. \ref{fig:S1}B albeit with inter-ciliary spacing of $3$ mm. \\

\textbf{Fabrication and surface treatment of molds.} The two mold parts are 3D-printed with VeroClear photoresin and SUP705 support material using a PolyJet Eden260VS printer (Stratasys). After printing, the molds are developed in a 5$\%$ KOH solution for $24$ hours to dissolve the support, then cleaned with a high-pressure waterjet. In particular, the bottom mold with outlet channels is placed in a KOH beaker under sonication to clear the fine channels forming the cilia. After cleaning, the molds are dried and their inner surfaces treated with mold release (Ease Release 200, Mann) to ensure smooth removal of cured silicone (Fig. \ref{fig:S1}D-E). Following each use, the molds undergo the same cycle of cleaning and mold-release treatment before reuse. \\

\textbf{Material preparation and injection procedure.} The overall procedure for preparation of silicone mixture and subsequently their injection molding is described in the following steps. First, the two components (A and B) of ED22F are loaded into a dual-compartment cartridge (MIXPAC EAAC400-01-10-01, 400 mL, 1:1 ratio). The cartridge then is mounted onto a pneumatic dispenser gun (MIXPAC DP400-85, Sulzer), where the components are mixed in-line during injection through a static mixer (QUADRO MFQ 05-24L, Sulzer) attached at the cartridge outlet. Second, the assembled 3D-printed molds are secured in a bench vise with their inlet holes oriented downward. The luer end of the static mixer is connected to a female luer (FTLL035), which in turn is coupled to the mold inlets via a $10$ cm silicone tube fitted with a male luer, thereby linking the mixer outlet to the mold inlet. Third, the prepolymer mixture is dispensed into a waste reservoir for several minutes to ensure that both the components are thoroughly mixed and air bubbles are purged. While maintaining a continuous flow, the tube outlet (male luer) are then inserted into the mold inlet to avoid introduction of stray bubbles (Fig. \ref{fig:S1}F-G). The silicone is injected at a controlled rate of $\sim 1.5$ mL/min, and filling the volume inside the molds ($\sim 7.5$ cm$^3$) lasts approximately $5$ minutes. Injection is maintained until silicone emerges from the mold outlets (corresponding to the ends of cilia), at which point a small pool of silicone is allowed to form above the upper inlets to prevent backflow (Fig. \ref{fig:S1}H). The silicone tube is finally disconnected under continuous flow, and the mold inlet is sealed using a rotating ring lock (FSLLR). \\

\textbf{Post processing.} After injection, the silicone is allowed to cure in the molds for $\sim 25$ minutes. The walkers is peeled out (Fig. \ref{fig:S1}I) and post-cured in an oven (UF30, Memmert) for $4$ hours at $60^\circ$C. After curing, both bidirectional and isotropic walkers weight $\sim 9$ g ($\sim 11$ g for the bi-rotational walker). The same injection and curing procedure is used for the enlarged single-cilium samples, with their respective scaled-up molds.

\subsection{Vibration platform} \label{section:vibration_platform}

\begin{figure}[p]
\centering
\includegraphics[width=\textwidth]{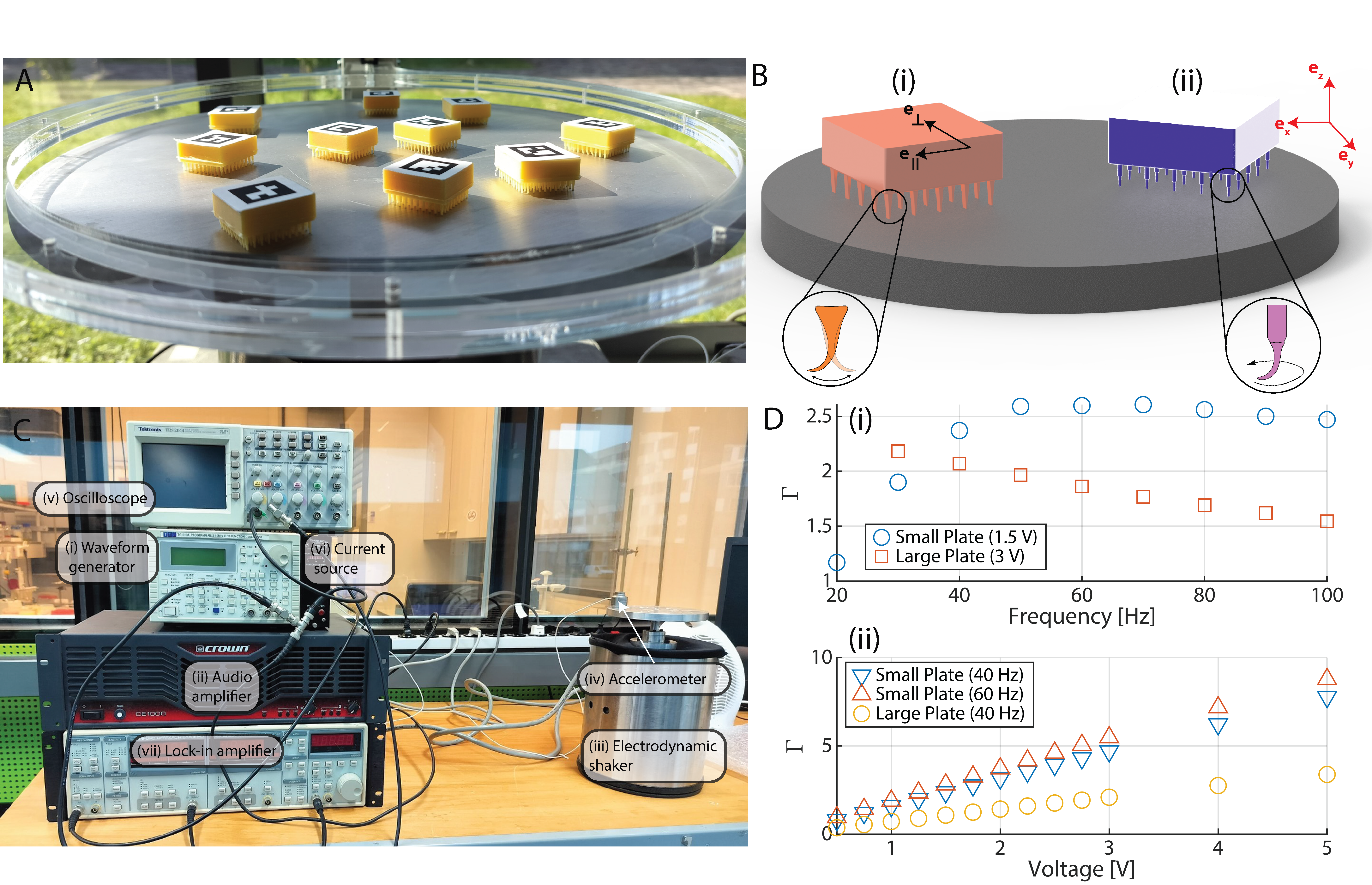}
\caption{Description of the vibration platform (see section \ref{section:vibration_platform}). (A) Photograph of multiple ciliary walkers placed on the vibrated aluminum plate. (B) Representative schematic showing (i) bidirectional and (ii) isotropic walkers, alongside the convention for the global frame of reference ($\boldsymbol{e}_x$, $\boldsymbol{e}_y$, $\boldsymbol{e}_z$) and the local frame of references for the walkers ($e_{\parallel}$, $e_{\perp}$). The direction $e_{\parallel}$ is picked along the printing direction of the 3D-printed molds of the walkers. (C) Complete experimental setup with all the hardware components for (i-iii) actuation and (iv-vii) calibration of vibrating plate. (D) Calibration curves of the two vibration plates used in experiments showing (i) frequency sweep at a fixed driving voltage and (ii) voltage sweep at $40$ Hz and $60$ Hz.}
\label{fig:S2}
\end{figure}

Figure \ref{fig:S2} illustrates the table-top experimental platform, together with multiple ciliary walkers placed on the vibrating substrate (Fig. \ref{fig:S2}A). Importantly, we define the global frame of reference of the lab ($\boldsymbol{e}_x$, $\boldsymbol{e}_y$, $\boldsymbol{e}_z$), and the local frames of reference of the walkers ($e_{\parallel}$, $e_{\perp}$), attached to body of the walkers (Fig. \ref{fig:S2}B). Below, we provide a detailed description of the different hardware components used in the vibration platform, as well as its operation and calibration procedure. \\

\textbf{Experimental setup.} We use an electrodynamic shaker (VTS systems VG100) as a our table-top vibration platform with a aluminum plate ($16$ cm diameter, mass $681$ g) mounted on it, which acts as the vibrating substrate for the walkers. The shaker is driven by a waveform generator whose signal is amplified via a commercial audio amplifier (Crown CE1000), with its gain fixed to $-14$ dB, and fed to the shaker. The electronic input via function generator (in volts) is transformed to a vertical acceleration (in m/s$^{2}$). The overall setup for vibrating the aluminum plate is described in Fig. \ref{fig:S2}C.i-iii. For experiments with multiple walkers, we use a larger aluminum plate ($36$ cm diameter, mass $2791$ g). \\

\textbf{Calibration of the platform.} Acceleration of the substrate is recorded using an accelerometer (Dytran 3120A, sensitivity $= 10$ mV/g) connected to a current source (Dytran 4119B, LIVM power unit), which converted the vibration signals to voltage. The output is fed to a lock-in amplifier (Stanford Instruments, SR830) together with the reference input from the waveform generator. The full setup is shown in Fig. \ref{fig:S2}C.iv--viii. To calibrate the platform, we measure how the electronic driving signal from the waveform generator and amplifier (in mV) translates to substrate acceleration (in m/s$^2$). We have performed two types of calibration sweeps: (i) acceleration versus frequency ($20$--$100$ Hz) at fixed input voltage, and (ii) acceleration versus input voltage ($0.5$--$3$ V) at fixed frequency. The calibration curves obtained from these sweeps are shown in Fig. \ref{fig:S2}D.i-ii.

\subsection{Instron measurements} \label{sec:instron}

We use a vertical uniaxial tensile testing machine (Instron, model 5965) to characterize the mechanical response of a single enlarged cilium under vertical compression and lateral shear. For both measurements, the machine is equipped with a static load cell (Instron 2530-100N) with a $5$ N capacity. The enlarged cilium is fabricated by injection molding (as described in Section \ref{sec:manufacturing}), using a dedicated single-cilium mold instead of the entire ciliary walker. The geometry is scaled by a factor of five relative to a single cilium of the bidirectional walkers (Fig. \ref{fig:S1}A-\emph{inset}), yielding a tapered profile (width = $6$ mm at the base and $1$ mm at the tip) and a rectangular cross-section (height = $30$ mm, depth = $7.5$ mm). \\

\begin{figure}[p]
\centering
\includegraphics[width=\textwidth]{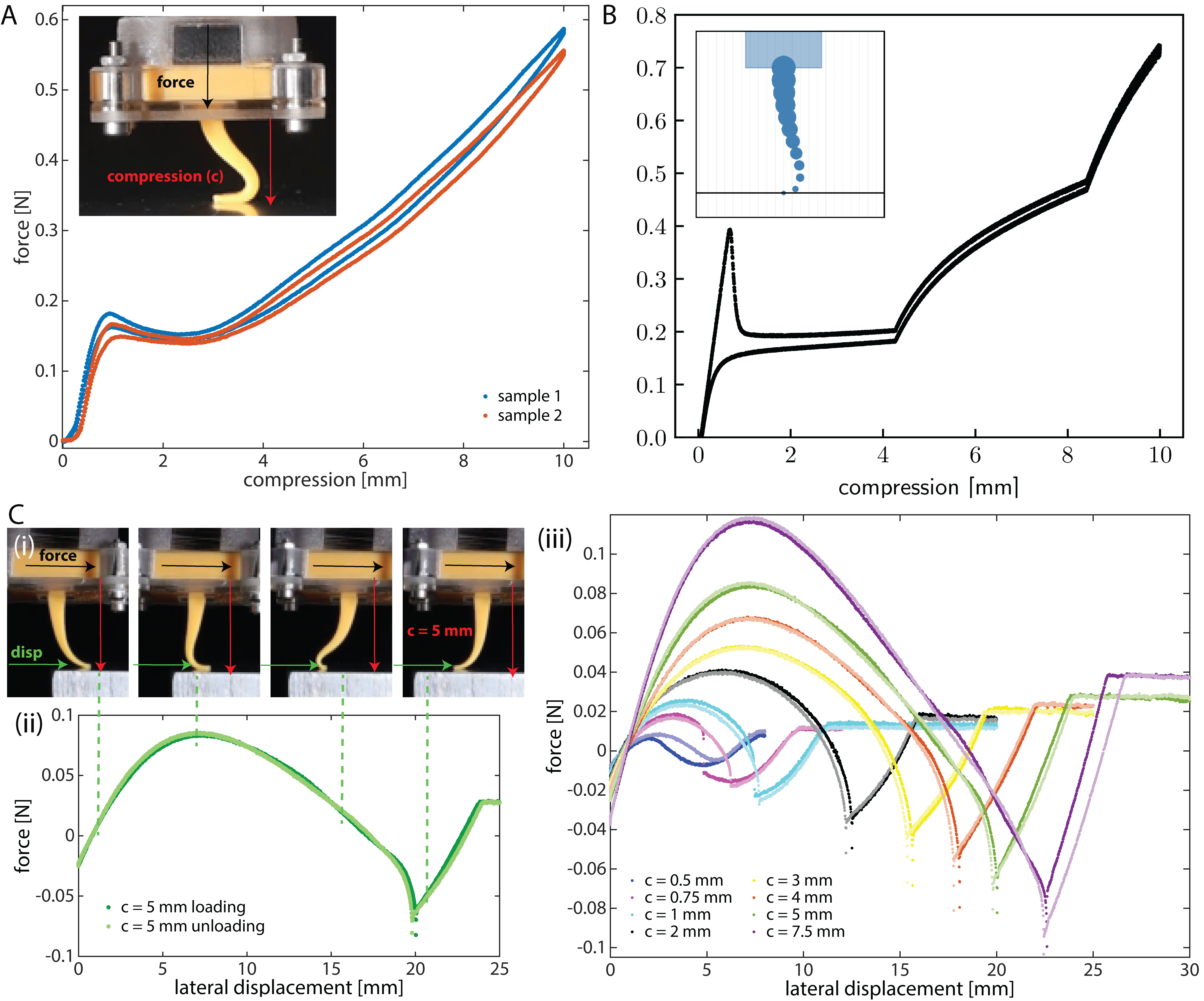}
\caption{Static characterization of a single cilium. (A) Force-displacement curve of a scaled-up elastic cilium (length = $30$ mm, see section \ref{sec:instron}) under vertical compression as obtained from experiments; and snapshot of the buckled cilia configuration at $c = 10$ mm (\emph{inset}). (B) Simulated counterpart for the single cilium model (see section \ref{sec:single_cilium}); and illustration of the buckled state at $c = 10$ mm (\emph{inset}). (C) (i) Time-lapse sequence of a cilium buckled under fixed compression (c, \emph{marked in red}) and laterally displaced (\emph{along green arrow}), highlighting bistability; and (ii) corresponding force-displacement curve showing the transitions between buckled states; (iii) force-displacement curves under varying fixed compressions (c). For all plots in (C), loading and unloading are represented by dark and lighter shades of the same color, respectively.}
\label{fig:S3}
\end{figure}

\textbf{Vertical compression of enlarged cilium (displacement controlled).} To characterize the response under compression, the enlarged cilium is mounted vertically on the lower clamp of the tensile-testing machine and compressed against a horizontal aluminum plate. The vertical displacement of the upper clamp is controlled while measuring the force exerted by the cilium along the compression axis (red and black arrows in Fig. \ref{fig:S3}A-\emph{inset}). Each sample is subjected to cyclic loading and unloading at a displacement rate of $5$ mm/s, with an sampling rate of $10$ Hz. A small hysteresis is observed between the loading and unloading curves, which we attribute to static friction at the contact with the aluminum plate and to viscous relaxation. However, we note that repeated cycles produce consistent and reproducible force-displacement curves.

The measured force-displacement curves exhibit the characteristic nonlinear behavior of a buckling column - an initial stiff regime, followed by softening upon buckling, and subsequent re-stiffening at large compression, corresponding to the positive-negative-positive slopes in the curve. Two independent samples are tested to account for fabrication variability (Fig. \ref{fig:S3}A). These measurements provide a reference for the buckling behavior of a single cilium, analogous to the deformation of ciliary walkers under external load, and are used below to calibrate the spring-mass single cilium model (Fig. \ref{fig:S3}B). \\

\textbf{Lateral shear of enlarged buckled cilium (displacement controlled).} To characterize the response of a pre-compressed cilium under lateral shear, the enlarged cilium is mounted horizontally on the upper clamp of the tensile-testing machine. An aluminum plate, oriented vertically at $90^{\circ}$ relative to the cilium, is fixed to the lower clamp (Fig. \ref{fig:S3}C-i). This L-shaped configuration effectively translates the vertical motion of the upper clamp into a controlled displacement of the cilium along the aluminum surface. Initially, the cilium is pre-compressed against the plate by a distance $c$ (denoted in red, Fig. \ref{fig:S3}C-i), which is varied systematically using a manual rotation stage (RP01/M, $\diameter$2", Thorlabs) to tune the pre-buckling condition (example shown for $c = 5$ mm).

In this configuration, the vertical displacement of the clamp produces a lateral shear motion between the cilium and the aluminum surface, while the corresponding shear force parallel to the surface (black arrow) is measured by the load cell. When the applied load exceeds a critical threshold, the cilium snaps between its two stable buckled states: left-leaning and right-leaning (along the green arrow in Fig.~\ref{fig:S3}C-i). Cyclic shear displacement cycles are applied at a rate of $10$ mm/s, and the resulting lateral force-displacement curves are recorded for different pre-compression levels, $c = 0.5-7.5$ mm (Fig. \ref{fig:S3}C-iii).

The measured curves exhibit a nonlinear response characteristic of the response of a bistable system to an external driving. Increasing pre-compression $c$ raises the \emph{energy barrier} separating the two buckled states. These measurements indicate that higher compression leads to larger shear forces at the transitions from one state to the other. This behavior provides insights into how the cilia of the walkers resist or switch their buckling orientation under different conditions.

\subsection{Image processing and data analysis}

The motion of ciliary walkers on the vibrating plate is analyzed across three complementary temporal and spatial scales. First, short-duration, high-speed videos of bidirectional walkers along a 1D track (Phantom v4.2 camera, $1000$ fps) to capture side-view dynamics during boundary collisions (approximately $5$ s of acquisition). Second, long-duration experiments of the different designs of ciliary walkers (GoPro Hero 10 camera, $30$ fps) to extract trajectories and characterize locomotion behaviors of walkers. Finally, we measure the statistics of the velocity and rotation rate to identify the cruise velocity $v_0$ and rotation rates $\omega_{0}^{\pm}$. Additionally, all demonstration videos provided in the Supplementary Movies (corresponding to Figs. 1 and 3 in the main text) are recorded (Canon EOS 850D camera, $50$ fps) for visualization purposes; these recordings were not used for quantitative analysis.

Below, we detail the image-processing and analysis pipeline for each stage, covering (i) segmentation of walkers in high-speed recordings, (ii) tracking of the motion of the walkers, and (iii) determination of the cruise velocity and rotation rates from the probability distributions.

\subsubsection{Segmentation and extraction of walker kinematics from high-speed movies (1D)} \label{sec:segmentation}

The high-speed recordings of bidirectional ciliary walkers are analyzed using a custom Python routine based on \texttt{OpenCV} and \texttt{scikit-image} to extract two key quantities per frame: the horizontal velocity of the walker ($V_x$, cm/sec) and the total contact area ($A$, mm$^2$) that cilia form with the vibrating plate. Each frame is first scaled-up (\texttt{cv2.resize}, $\times$10), and two reference points at the ends of the oscillating plate are identified via thresholding (\texttt{cv2.threshold}) to define a dynamic region of interest (ROI) containing the walker. Frames are rotated to align the plate horizontally and cropped to isolate the walker.

Within the ROI, the rectangular substrate that forms body of walker is detected. Frames are converted to grayscale, subjected to contrast enhancement (\texttt{cv2.createCLAHE}), thresholding (\texttt{cv2.threshold}), to extract contours (\texttt{cv2.findContours}) of this substrate. The lower boundary of the substrate, where cilia are connected, defines the base of the rectangle, from which the mean horizontal position (center of mass, COM) is computed. Cilia are segmented by refining the binarized ROI with morphological operations (\texttt{cv2.erode}, \texttt{cv2.dilate}). The midlines of each cilium are skeletonized (\texttt{skimage.morphology.skeletonize}), and connected components (\texttt{cv2.connectedComponentsWithStats}) shorter than a minimum threshold are removed.

Contact points with the substrate are identified as the lowest $y$-coordinate pixels in each skeletonized cilium, merged along $x$-direction to compute total projected length on the plate. This length is then multiplied by a fixed cilium width ($1.5$ mm) to yield the contact area $A$. Time series of COM positions and contact areas were smoothed (\texttt{scipy.signal.savgol filter}), with horizontal velocity, $V_x$, derived from the framewise derivative of COM.

These quantities capture the locomotion dynamics of the walkers. High-speed recordings in the \emph{reversals} regime are analyzed using this pipeline to characterize one full sequence of the walker translating to the nearest boundary and reversing its locomotion. Each locomotion mode is characterized by a work-generating limit cycle, with $A$ and $V_x$ oscillating approximately in quadrature (Fig. \ref{fig:S4}A-i,ii). The mean translational velocity of the walker during these cycles, denoted as $v_0 = \langle V_x \rangle$, corresponds to the cruise velocity reported in the main text. Upon autonomous reversal at boundaries, the walker transitions smoothly between distinct limit cycles (Fig. \ref{fig:S4}A-iii, light-to-dark blue).

\subsubsection{Tracking motion of ciliary walkers}

Top-view recordings of ciliary walkers carrying printed ArUco markers (mounted on top weights) are analyzed using a custom Python routine based on \texttt{OpenCV} with post processing in MATLAB. Each frame is cropped to a fixed region of interest (ROI), converted to grayscale, and contrast-enhanced using adaptive histogram equalization (\texttt{cv2.createCLAHE}). ArUco markers were detected with \texttt{cv2.aruco.detectMarkers} (\texttt{DICT\_4$\times$4\_50}), providing the centroid position $(x, y)$ and body orientation $\theta$. The instantaneous translational velocity $(V_x, V_y)$ and angular velocity $\omega = \mathrm{d}\theta/\mathrm{d}t$ are obtained from framewise displacements and exported files for post-processing. The velocity components are smoothed using a Gaussian kernel and transformed into the local body frame defined by the orientation vectors obtained from ArUco markers $(\boldsymbol{e}_{\parallel}, \boldsymbol{e}_{\perp})$, yielding the parallel and perpendicular components $(V_{\parallel}, V_{\perp})$. For 1D locomotion, only the axial component $V_{\parallel}$ is retained for time-series and distribution analyses, while for 2D locomotion, the full set $(V_{\parallel}, V_{\perp}, \omega)$ is used to construct velocity histograms and perform subsequent statistical analyses.

\subsubsection{Analysis of the probability distributions} \label{sec:peak_determination} 

In 2D experiments, the cruise velocity $v_0$ and rotation rates $\omega_0^{\pm}$, if they exist, are determined by extracting the positions of the peaks in the distributions of $|\boldsymbol{V}|$ and $\omega$. This procedure is represented in Fig. \ref{fig:S10} for a characteristic example. We first represent the probability distribution of ($V_{\parallel}$, $V_{\perp}$,$\omega$) in the plane $|\boldsymbol{V}|-\omega$, where three lobes are clearly visible (Fig. \ref{fig:S10}A). They are respectively associated with the translational modes and with the two rotational modes with opposite handedness. Then, we plot the probability distribution of $|V|$ (Fig. \ref{fig:S10}B). In this case, we find two peaks: one at small velocity associated with the two rotational lobes, and one at larger velocity associated with the translational lobe. The latter maxima is used to define the cruise velocity $v_0$. Eventually, we plot the probability distribution of $\omega$ (Fig. \ref{fig:S10}C), where we can clearly see the two rotational lobes with finite and almost opposite rotation rates. These two maxima are used to define $\omega_0^{\pm}$. We note that we use gaussian filtering to reduce the fluctuations and the noise in the distributions (and then determine more accurately the position of the peaks), but with a small enough window to avoid changing quantitatively the shape of the peaks.

\newpage

\section{Modeling of a single vibrated cilium} \label{sec:single_cilium}

In this section, we describe the simulation framework and parameters to model a single vibrated cilium. We start with approximating a single cilium of our bidirectional walkers as a lumped-parameter (mass-spring) model. Each cilium is modeled as a series of discrete nodes connected by linear axial springs representing axial tension/compression stiffness, linear torsional (bending) springs at each node to capture bending resistance, and a lumped mass at each node to capture inertial effects comparable to the substrate supported by the cilia in our walker. We divide the cilia with \(N = 12\) equally spaced \emph{nodes}, giving \((N-1)\) \emph{segments}. 
Thereon, we describe the geometry and material properties of the cilium. These estimates allow us to calibrate the axial and torsional spring constants and the lumped nodal mass of the cilium. 

\subsection{Discretization}

The cilium of length $L$ is divided into \( N-1 \) segments, with \( N \) nodes, such that each segment has length $L_{\text{segment}} = L/(N-1)$.
To approximate the varying bending stiffness along the tapered length of the cilium, we evaluate the geometric properties at the \emph{midpoints} of each segment.

\[
  M_{\text{total}} = \rho \int_0^L A(z)\,\mathrm{d}z
  \;\approx\; \rho \sum_{i=1}^{N-1} A(z_{k_s,i})\,L_{\text{segment}},
\]
where \(A(z_{k_s,i})\) is the local average cross-sectional area of the \(i\)th segment evaluated at its midpoint \(z_{k_s,i} = (i - \tfrac{1}{2})L_{\text{segment}}\).  
Each segment mass is then distributed to its adjacent nodes, resulting in nodal masses \(m_i\) located at node positions
\[
  z_{m,i} = i\,L_{\text{segment}}, \qquad i = 0,1,\dots,N-1.
\]
The total mass is thus distributed in proportion to the local segment area, giving
\[
  m_i = M_{\text{total}}
  \frac{A(z_{k_s,i})\,L_{\text{segment}}}
       {\sum_{j=1}^{N-1} A(z_{k_s,j})\,L_{\text{segment}}},
  \qquad
  \sum_i m_i = M_{\text{total}}.
\]
This produces larger masses near the wider base and smaller masses near the tapered tip, consistent with the geometry.

\subsection{Geometry and material properties of the cilium}

The tapered profile of the rectangular-section cilium has the following properties: total length: \( L \), base width: \( w_{\text{base}} \), tip width: \( w_{\text{tip}} \), constant thickness (out-of-plane): \( t \), Young's modulus: \( E \) and density: \(\rho\). The width varies linearly along the length of the cilium, according to:
\begin{equation} \label{eq:evolution_width}
    w(z) = w_{\text{base}} + \left( \frac{w_{\text{tip}} - w_{\text{base}}}{L} \right) z,
\end{equation}
where \( z \) is the distance from the base (\( z=0 \)) to the tip (\( z=L \)).

\subsection{Description of spring constants and nodal mass}
This section entails the calculation of spring stiffness for axial $k_s$ and torsional $k_b$ springs that describe each segment and the mass $m$ of each nodal segment.

\subsubsection{Nodal mass}
Assume the cilium is of uniform density \(\rho\). Its total mass is
\[
  M_{\text{total}} \;=\; \rho \, A \, L.
\]
where $A = (1/L)\int A(z_i)$ is the mean cross sectional area, with $A(z_i)$ the cross sectional area of segment $i$. 
where \(A(z_i)\) is the local cross-sectional area of segment \(i\) evaluated at its midpoint \(z_i = (i - \tfrac{1}{2})L_{\text{segment}}\).  
In the discrete model, this total mass is distributed among the segments in proportion to their local cross-sectional area, such that
\[
  m_i = M_{\text{total}}
  \frac{A(z_i)\,L_{\text{segment}}}
       {\sum_{j=1}^{N-1} A(z_j)\,L_{\text{segment}}},
  \qquad
  \sum_i m_i = M_{\text{total}}.
\]
This yields heavier masses near the base, where the cilium is wider, and lighter masses near the tip.  

\subsubsection{Axial stiffness}
Consider each segment of cilium behaving like a 1D axial spring. The axial (tensile/compressive) stiffness of the i$^{th}$-segment is:
\begin{equation} \label{eq:linear_spring}
    k_s^{(i)} = \frac{E \, A(z_{k_s,i})}{L_{\text{segment}}}
    = (N-1) \,\frac{E \, A(z_{k_s,i})}{L},
\end{equation}
The cross-sectional area at position \(z_{k_s,i}\) is given by:
\begin{equation}
    A(z_{k_s,i}) = 
    t \left[
      w_{\text{base}} +
      \left( \frac{w_{\text{tip}} - w_{\text{base}}}{L} \right)
      z_{k_s,i}
    \right],
\end{equation}

\subsubsection{Torsional stiffness}
The bening stiffness of the cilium is modeled with a discrete torsional springs at the node connecting two segments. To compute the torsional stiffness, we need to evaluate the moment of inertia, the bending moment and the curvature. First, we compute the \emph{second moment of area} \( I(z) \) at each node of the cilium:
\begin{equation}
    I(z) = \frac{1}{12} t \cdot w(z)^3.
\end{equation}
Substituting \( w(z) \) with Eq. (\ref{eq:evolution_width}) yields:
\begin{equation}
    I(z_{k_b,i}) = \frac{1}{12} t \left( w_{\text{base}} + \frac{(w_{\text{tip}} - w_{\text{base}})}{L} z_{k_b,i} \right)^3.
\end{equation}
From Euler-Bernoulli beam theory for small deflections, the relationship between bending moment \(M\) and curvature \(\kappa\) is:
\[
  M = E\, I \,\kappa, 
  \quad \text{with} \quad \kappa \approx \frac{\Delta \theta}{L_{\text{segment}}}
  \quad (\text{for small rotations } \Delta\theta).
\]
Thus, at each node, the moment can be approximated by
\[
  M \;=\; E \, I(z_{k_b,i}) \,\frac{\Delta\theta}{L_{\text{segment}}}.
\]
Since \(\Delta\theta\) is the rotation at node $i$, the torsional (bending) spring constant (torque per unit angle) is:
\[
  k_b^{(i)} \;=\; \frac{M}{\Delta\theta}
         \;=\; \frac{E \, I(z_{k_b,i})}{L_{\text{segment}}}.
\]

\subsection{Static response of cilium} 
To replicate the nonlinear response of a buckled cilium, we calibrate the spring network by fitting its force-displacement curve to that of a vertically compressed, scaled-up elastic cilium (Fig. \ref{fig:S3}A-B). The calibrated model is finally placed in a simulation environment with tunable vibration amplitude $\Gamma$ and load $W$ (SI Movie 3).

\subsection{Simulation framework and dynamic response of the cilium}\label{sec:single_cilium:dynamics}

To capture the behavior of the cilium under vertical vibration, we numerically integrate the equations of motion derived from the lumped-parameter formulation. Each node \(i \in [1, N]\) has a two-dimensional position vector $\boldsymbol{r}_i(t) = \left[ x_i(t), z_i(t)\right]$ and velocity $\dot{\boldsymbol{r}}_i(t) = \left[ \dot{x}_i(t), \dot{z}_i(t)\right]$. The full system displacement vector is written as $\boldsymbol{u} = \left[ \boldsymbol{r}_1 (t), \boldsymbol{r}_2 (t), \dots, \boldsymbol{r}_N (t) \right] \in \mathbb{R}^{2N}$. The force balance at each node follows from Newton's law:
\begin{equation}
\boldsymbol{M} \ddot{\boldsymbol{u}}(t) = 
\boldsymbol{F}_{\text{lin}}(\boldsymbol{u})
+ \boldsymbol{F}_{\text{tor}}(\boldsymbol{u})
+ \boldsymbol{F}_{\text{fric}}(\boldsymbol{u}, \dot{\boldsymbol{u}}, t) + \boldsymbol{F}_{\text{grav}} 
- c_{\text{damp}} \dot{\boldsymbol{u}},
\label{eq:motion}
\end{equation}
where \( \boldsymbol{M} \) is a diagonal mass matrix containing the lumped nodal masses, \( \boldsymbol{F}_{\text{lin}} \) are the elastic forces from the linear (axial) springs, \( \boldsymbol{F}_{\text{tor}} \) are the bending forces from the torsional springs, \( \boldsymbol{F}_{\text{grav}} \) are the gravity forces, \( \boldsymbol{F}_{\text{fric}} \) accounts for ground contact friction, and \(c_{\text{damp}}\) is a viscous damping coefficient.

\subsubsection{Linear spring forces}
For each pair of connected nodes \((i, j)\), the relative displacement is
\[
\boldsymbol{d}_{ij} = \boldsymbol{r}_j - \boldsymbol{r}_i,
\qquad
\delta l_{ij} = \lVert \boldsymbol{d}_{ij} \rVert - L_0,
\]
where \( L_0 \) is the segment rest length. The linear elastic force acting on node \(i\) due to segment \((i, j)\) is
\begin{equation}
\boldsymbol{f}_{ij}^{\text{lin}} = 
- k_{s}^{(i)} \, \delta l_{ij} \, 
\frac{\boldsymbol{d}_{ij}}{\lVert \boldsymbol{d}_{ij} \rVert},
\end{equation}
with stiffness \(k_{s}^{(i)}\) defined previously in Eq.~(\ref{eq:linear_spring}). These forces are equal and opposite on the adjacent nodes.

\subsubsection{Torsional spring forces}
Bending resistance between three consecutive nodes \((i-1, i, i+1)\) is modeled as a torsional spring at node \(i\). The instantaneous bending angle is defined as
$\theta_i = \arctan \!\big(
(\boldsymbol{r}_i - \boldsymbol{r}_{i-1}) \times (\boldsymbol{r}_i - \boldsymbol{r}_{i+1}),
\, (\boldsymbol{r}_i - \boldsymbol{r}_{i-1}) \cdot (\boldsymbol{r}_i - \boldsymbol{r}_{i+1}) \big)$ and the corresponding bending torque is 
\begin{equation}
\tau_i = -k_b^{(i)} \, (\theta_i - \pi),
\end{equation}
which produces a distributed nodal force vector proportional to the spatial derivatives of \(\theta_i\) with respect to each node position. This effectively penalizes local curvature and enforces elastic resistance to bending deformations.

\subsubsection{Friction, contact, and gravity}

Each node can interact with a vertically oscillating ground described by $z_{\text{ground}}(t) = A \sin\!\left( 2\pi f t \right) + z_0$, where \(f\) is the vibration frequency and \(z_0\) is a positive vertical offset. The vibration amplitude \(A\) is defined in terms of the dimensionless acceleration \(\Gamma\) as described in the main text. In addition to the elastic forces, each node \(i\) is subjected to a vertical force which consists of gravity, and a ground reaction when the node penetrates the vibrating plate:
\begin{equation}
F_{z,i} =
\begin{cases}
k_N \, (z_{\text{ground}}(t) - z_i), & z_i < z_{\text{ground}}(t),\\[4pt]
0, & z_i \ge z_{\text{ground}}(t),
\end{cases}
\end{equation}
where \(k_N\) is an effective ground stiffness parameter. Moreover, an horizontal friction force acts on node $i$ when it is in contact with the substrate (\(z_i < z_{\text{ground}}(t)\)). This tangential contact friction is modeled using a smooth exponential regularization of Coulomb's law:
\begin{equation}
F_{x,i} = 
- \mu_g \, F_{N,i} \,
\big( 1 - e^{-10 |\dot{x}_i|} \big)
\, \mathrm{sgn}(\dot{x}_i),
\end{equation}
where \(F_{N,i}\) is the magnitude of the normal reaction from the ground, and \(\mu_g\) is a friction coefficient.  
This nonlinear form captures the transition from static to dynamic friction smoothly as the slip velocity increases.
The total ground reaction on each node is therefore
\[
\boldsymbol{F}_{\text{fric},i} = 
\begin{bmatrix}
F_{x,i} \\[4pt]
F_{z,i}
\end{bmatrix}.
\]
Eventually, the gravity force is constant on each node and can be written as \(-m_i g \, \hat{\boldsymbol{z}}\).

\subsubsection{Boundary conditions and base constraint}
The first (top) node is constrained in both coordinates to enforce a fixed base segment orientation. At every time step, we impose $x_0(t) = x_1(t)$ and $z_0(t) = z_1(t) + L_0$ and the corresponding velocity constraint $\dot{\boldsymbol{r}}_0(t) = \dot{\boldsymbol{r}}_1(t)$.
These constraints keep the base segment \((0,1)\) strictly vertical (fixed orientation) with constant separation \(L_0\), while letting the chain deform downstream. In practice, this constraint acts as a clamped-like boundary condition that fixes the angle of the first segment relative to the ground, ensuring no rotation of the \((0,1)\)-link.

To confine motion inside the track and model the interaction with the ends of the vibrating plate, we include interactions with two vertical boundaries positioned at $x = \pm L_{\rm{track}}/2$, with $L_{\rm{track}} = 20$ cm. The second node (index \(i=1\)) experiences an additional horizontal reaction force that pushes it back into the track whenever it penetrates inside the vertical walls. The force acts purely along the \(x\)-axis, with magnitude proportional to a reference vertical offset, and is given by
\begin{equation}
F^{\text{wall}}_{x,1} =
\begin{cases}
\;\;k_W \, \big| x_1 + L_{\rm{track}}/2 \big|, & x_1 < - L_{\rm{track}}/2, \\[6pt]
-\,k_W \, \big| x_1 - L_{\rm{track}}/2 \big|, & x_1 > L_{\rm{track}}/2, \\[6pt]
0, & \text{otherwise},
\end{cases}
\qquad
F^{\text{wall}}_{z,1} = 0,
\label{eq:wall_force}
\end{equation}
where \(k_W\) is the wall reaction coefficient is a fixed vertical reference. This penalty-like reaction provides a lateral restoring force that depends on the node's vertical position and only activates upon wall penetration, leaving all other nodes unaffected. In the simulations, this term is used to emulate confinement by vertical guides without introducing additional vertical loads.

\subsubsection{Compression test simulations}
A quasi-static vertical compression test, analogous to an Instron experiment, is performed on the scaled cilium. The bottom node is fixed, while the top node is driven vertically downward at a constant velocity of $v_y = -0.072~\mathrm{cm/s}$ for a total duration of $5~\mathrm{s}$, corresponding to a $60\%$ axial compression. Horizontal motion of both end nodes is constrained to ensure purely axial loading. We find a qualitatively similar mechanical response to vertical compression in simulations and in the experimental scaled-up cilium (Figs. \ref{fig:S3}A-B). Moreover the configuration of the buckled simulated and experimental cilia are very close, with a strong deformation and a large contact area with the plate (Figs. \ref{fig:S3}A-B, \textit{insets}).

\subsubsection{Damping}
A small viscous damping term \( -c_{\text{damp}} \dot{\boldsymbol{u}} \) is added, with \(c_{\text{damp}} = 10^{-3}\). 

\subsubsection{Numerical integration}
The system of \(4N\) first-order differential equations corresponding to Eq.~(\ref{eq:motion}) is integrated using the explicit Runge-Kutta method (RK45) from the \texttt{scipy.integrate.solve\_ivp} library (timestep: $\Delta t = 10^{-3}\,\mathrm{s}$, tolerances: $\text{rtol} = 10^{-8}, \text{atol} = 10^{-10}$). The total simulation duration is \(T_{\text{total}} = 20\,\mathrm{s}\).

\subsubsection{Parameter summary}
The key simulation parameters are summarized as:
\[
E = 1.2\times10^6~\mathrm{N/m^2}, \quad
L = 6\times10^{-3}~\mathrm{m}, \quad
N = 12, \quad
g = 9.8~\mathrm{m/s^2}.
\]
The resulting framework provides a physically grounded simulation of a single cilium subjected to vertical vibration.

\begin{table}[h!]
\centering
\begin{tabular}{llll}
\hline
\textbf{Category} & \textbf{Symbol / Name} & \textbf{Value} & \textbf{Units / Description} \\
\hline
\textbf{Material} & $E$ & $1.2\times10^{6}$ & N/m$^{2}$ (Young's modulus) \\
 & $W_{\text{total}}$ & $2.0\times10^{-4}$ & kg (Total weight of cilium) \\
 & $W_{\text{body}}$ & $8.2\times10^{-3}$ & kg (Weight of one cilium body) \\
\hline
\textbf{Simulation} & $\Delta t$ & 0.001 & s (Time step) \\
 & $T_{\text{total}}$ & 0.25 & s (Total simulation time) \\
 & $f$ & 40 & Hz (Vibration frequency) \\
 & $g$ & 9.8 & m/s$^{2}$ (Gravitational acceleration) \\
 & $\mu_g$ & 5.0 & -- (Ground friction coefficient) \\
 & $F_N$ & 200 & N (Ground normal force) \\
 & $k_W$ & 150 & N/m (Wall reaction constant) \\
 & walls & True & (Wall interaction flag) \\
 & instron\_test & True & (Static test flag) \\
\hline
\textbf{Solver} & method & RK45 & (Integration method) \\
 & rtol & $1\times10^{-8}$ & (Relative tolerance) \\
 & atol & $1\times10^{-10}$ & (Absolute tolerance) \\
\hline
\end{tabular}
\caption{Summary of simulation and material parameters for the single cilium model.}
\end{table}

\newpage
\section{Simulations of underdamped rigid active solids} \label{sec:active_solid} 

Here, we describe the general equations and the numerical scheme used to simulate the dynamics of the rigid active solids discussed in the main text. 

We consider a ciliary walker as a rigid 2D square, embedded with $N$ freely-rotating active units which represent the propulsion forces generated by each cilia \cite{baconnier2022selective}. In this context, the motion of the active structure originates from the collective configuration of the active forces. Moreover, we consider that each active unit reorients toward its velocity, a generic process referred to as self-alignment \cite{baconnier2025self}. The thorough derivation of the rigid limit can be found in \cite{hernandez2024model}. Below, we summarize the main steps to go from the agent-based model to the equations for the rigid body motions.

\subsection{General rigid equations} We start from the most general equations of motion for a single active unit embedded in a spring network. The position $\boldsymbol{r}_i$ and orientation $\boldsymbol{\hat{n}}_i$ of active particle $i$ evolve according to \cite{baconnier2022selective}:
\begin{subequations} \label{eq:app:general_1}
\begin{align}
m \frac{d^2 \boldsymbol{r}_i }{dt^2} &= - \gamma \frac{d \boldsymbol{r}_i }{dt} + F_0 \boldsymbol{\hat{n}_i} - k \sum_{j \in \partial i} (| \boldsymbol{r}_i - \boldsymbol{r}_j | - l_0) \boldsymbol{\hat{e}}_{ij}, \label{eq1:app:general_1} \\
 \frac{d \boldsymbol{\hat{n}}_i }{dt} &= \frac{1}{l_a} (\boldsymbol{\hat{n}}_i \times \frac{d \boldsymbol{r}_i }{dt}) \times \boldsymbol{\hat{n}}_i + \sqrt{2D_{\theta}} \xi_i \boldsymbol{\hat{n}}_i^{\perp}, \label{eq2:app:general_1}
\end{align}
\end{subequations}
where each active unit has a mass $m$ and exerts a force $F_0$ in the direction of its polarity $\boldsymbol{\hat{n}}_i$, where $k$ and $l_0$ are the springs' stiffness and rest length, and where $\gamma$ is an effective friction coefficient. In the absence of confinement (and without springs, i.e., freely-moving in isolation), a single active unit thus moves with a cruise velocity $v_0 = F_0/\gamma$. The orientation dynamics Eq. (\ref{eq2:app:general_1}) contains the key ingredient, specific to the model, namely the presence of a self-aligning torque of the orientation $\boldsymbol{\hat{n}}_i$ towards the velocity $\boldsymbol{v}_i$. This torque originates from the fact that the dissipative force is not symmetric with respect to the propulsion direction $\boldsymbol{\hat{n}}_i$ when $\boldsymbol{v}_i$ is not aligned with $\boldsymbol{\hat{n}}_i$ \cite{baconnier2025self}. Importantly, the self-alignment torque is proportional to the velocity, giving rise to an alignment length $l_a$ \cite{baconnier2025self}.  Finally, the orientation dynamics contains a delta-correlated Gaussian noise $\xi_i(t)$ with zero mean and correlations $\langle \xi_i(t) \xi_j(t') \rangle = \delta(t - t') \delta_{ij}$; and $D_{\theta}$ is the rotational diffusion coefficient. Those orientation fluctuations are not of thermal origin but model the mechanical noise present in the experiments.

The rigid limit corresponds to the limit where $k \rightarrow +\infty$, all other parameters being held constant. In this case, only a subsets of the normal modes of the elastic structure can be actuated by the active dynamics: those which do not deform any bonds, i.e., the \textit{zero modes}. We denote $\mathcal{F}$ the subset of zero modes (which, for a mechanically stable structure in free boundary conditions in the 2D plane, corresponds to the two rigid body translations, say along $\boldsymbol{\hat{e}}_x$ and $\boldsymbol{\hat{e}}_y$, and to one rigid body rotation). Choosing the units of time and length such that $r_0 = l_0$ and $t_0 = l_0/v_0$, Eqs. (\ref{eq:app:general_1}) can be re-cast into \cite{hernandez2024model}:
\begin{subequations} \label{eq:app:rigid_1}
\begin{align}
M \frac{d^2 \boldsymbol{r}_i }{dt^2} +  \frac{d \boldsymbol{r}_i }{dt} &= \sum_{q \in \mathcal{F}} \langle \boldsymbol{\varphi}^q | \boldsymbol{\hat{n}} \rangle \boldsymbol{\varphi}^q_i, \label{eq1:app:rigid_1} \\
 \frac{d \theta_i }{dt} &= - \frac{1}{\tau_n} \frac{\partial V}{\partial \theta_i} + \sqrt{\frac{2D}{\tau_n}} \xi_i, \label{eq2:app:rigid_1}
\end{align}
\end{subequations}
where $\theta_i$ is the angle of $\boldsymbol{\hat{n}}_i$ with respect to the $x$-axis, $M = m v_0/\gamma l_0$, $D = D_{\theta} l_a / v_0$, $\tau_n = l_a /l_0$, and with:
\begin{equation}
V( \{\boldsymbol{\hat{n}}_j \} ) = - \frac{1}{2} \sum_{q \in \mathcal{F}} \langle \boldsymbol{\varphi}^q | \boldsymbol{\hat{n}} \rangle^2,
\end{equation}
where $\boldsymbol{\varphi}^q_i$ is the displacement vector associated with particle $i$ in the $q$-th zero mode. Altogether, the dynamics is determined by the set of zero modes (which depend on dimensionality and on the geometry of the rigid structure), and by $3$ dimensionless parameters: $M$ (dimensionless inertia; the ratio of linear momentum to frictional losses), $\tau_n$ (strength of self-alignment), and $D$ (amplitude of angular noise).

Without inertia ($M = 0$) and for small enough noise ($D < D^{\star} = 0.5$), it was shown that such models give rise to multiple modes of locomotion associated with different steady motions along the zero modes (rotational and translational motions in the plane) \cite{hernandez2024model}.

In the following, we simulate a $10 \times 10$ rigid square lattice of active forces, keeping $\tau_n = 0.5$ fixed (the active forces reorient over half the inter-cilia distance), which appears to be close to the typical experimental value. Moreover, consistently with experiments, we consider noise amplitudes small as compared to the critical noise amplitude at the order/disorder transition, i.e., $D \ll D^{\star} = 0.5$. 

\subsection{Dynamics in a 1D track}\label{sec:active_solid:standard} In a 1D track (say, along $\boldsymbol{\hat{e}}_x$), the elastic structure has a single zero mode, associated to the rigid body translation along the track ($\boldsymbol{\varphi}^q_i = \boldsymbol{\hat{e}}_x/\sqrt{N}$). Between two collisions with the edges of the track, the center of mass $X_c$ and the orientations $\theta_i$ evolve following:
\begin{subequations} \label{eq:app:rigid_1D}
\begin{align}
M \frac{d^2 X_c }{dt^2} +  \frac{d X_c }{dt} &= \frac{1}{N} \sum_{i} \cos \theta_i, \label{eq1:app:rigid_1D} \\
 \frac{d \theta_i }{dt} &= - \frac{1}{\tau_n} \frac{d X_c}{d t} \sin \theta_i + \sqrt{\frac{2D}{\tau_n}} \xi_i. \label{eq2:app:rigid_1D}
\end{align}
\end{subequations}
Moreover, we consider perfectly elastic collisions at the walls, positioned at $x = \pm L_{\textrm{track}}/2$, such that the velocity $dX_c /dt$ of the rigid active solid is reversed when a collision occurs. Finally, in the initial condition, the active solid is at rest in the center of the track, and the orientations of the active units are drawn at random (Fig. \ref{fig:S7}A).

We explore the dynamics of the velocity and the velocity distribution in simulations with a small noise $D = 10^{-3}$. First, we find that the velocity $V_x$ converges exponentially toward the values $V_x \simeq \pm 1$ depending on the initial condition. This convergence happens over a characteristic timescale which is set by the inertia $M$. 

Remarkably, different inertia $M$ also lead to different behaviors upon collisions. For small inertia $M$ (Fig. \ref{fig:S7}E-G), after the first collision, it takes very little distance (small as compared to $l_a$) for the active forces to restore motion back in the original direction (toward the wall). This leads to a second smaller collision (with a smaller velocity), and even to many subsequent smaller and smaller collisions, with the active forces eventually frozen in the direction of the wall: the active solid is \textit{trapped}. Note that the motion can eventually resume in the other direction at long times when angular diffusion of the polarities lead the polarization $P_x$ to reverse through noise. In contrast, for larger inertia $M$ (Fig. \ref{fig:S7}B-D), the active forces polarize in the direction opposite from the wall after a collision over a characteristic distance $l_a$: the active solid autonomously reverses direction after the collision. The transition from \textit{trapped} to \textit{reversals} as $M$ increases is well captured by the probabilities of the trapped state $|V_x| \sim 0$ (Fig. \ref{fig:S7}H), and happens around $M \simeq 6$.
\subsection{Dynamics of bidirectional active solids in a 1D track}\label{sec:active_solid:bidirectional}
In this section, we introduce a model for bidirectional ciliary walkers that is more faithful to the experiments than the model above. Indeed, Eqs. (\ref{eq:app:rigid_1D}) assume that the active forces takes the form of polar unit vectors in the plane. Instead, in experiments with bidirectional ciliary walkers, the active forces live in the direction of the 1D track, and they have no contribution perpendicular to the track. In the following, we consider that active forces are described by a polarization $m_i$ taking values in $\left[ -1, 1 \right]$, and whose dynamics follow from:
\begin{subequations} \label{eq:app:rigid_1D_bidirectional}
\begin{align}
M \frac{d^2 X_c }{dt^2} +  \frac{d X_c }{dt} &= \frac{1}{N} \sum_{i} m_i, \label{eq1:app:rigid_1D_bidirectional} \\
 \frac{d m_i }{dt} &= \frac{1}{\tau_n} \frac{d X_c}{d t} (1 - m_i^2) + \sqrt{\frac{2D}{\tau_n}} \xi_i. \label{eq2:app:rigid_1D_bidirectional}
\end{align}
\end{subequations}
where $X_c$ is the x-position of the center of mass of the rigid active solid. Moreover, we again consider perfectly elastic collisions at the walls, positioned at $\pm L_{\rm track}/2$, such that the velocity of the rigid active solid is reversed when a collision happens. Finally, in the initial condition, the active solid is at rest in the center of the track, and the polarizations $m_i$ of the active units are drawn are flatly drawn in $\left[ -1, 1\right]$ (Fig. \ref{fig:S8}A).

We explore the dynamics of the velocity and the velocity distribution in simulations. We find the same behavior as in the previous section. For small inertia (Fig.
\ref{fig:S8}-left, bottom), the active solid is trapped, even though motion can resume in the other direction through thermal fluctuations. In contrast, for larger inertia (Fig. \ref{fig:S8}-left, top), the active solid autonomously reverses direction after the collision. The sharp transition from trapped to reversing as $M$ increases is well captured by the probability of the trapped state $Vx \simeq 0$ (Fig. \ref{fig:S8}-right), and happens around $M \simeq 4$.

Altogether, we find that the inertia-controlled transition from trapped to reversals is common to both models of rigid active solids in a 1D track. This is expected given the two models share the same ingredients and only differ in the details of the reorientation dynamics of the active units (see Eqs. (\ref{eq2:app:rigid_1D}) and (\ref{eq2:app:rigid_1D_bidirectional})).
\subsection{Dynamics in a 2D arena}\label{sec:active_solid:2d} Finally, we come back to the model with polar active forces, i.e., Eqs. (\ref{eq:app:rigid_1}). In a 2D arena, the elastic structure has three zero modes, associated to the translational and rotational rigid body motions in the plane. In this case, we have to describe the dynamics of the $x$-and $y$-components of the center of mass, as well as of the angle $\phi$ of the structure in the reference frame of the lab:
\begin{subequations}
\label{eq:app:rigid_2D}
\begin{align}
M \frac{d^2 X_c }{dt^2} +  \frac{d X_c }{dt} &= \frac{1}{N} \sum_{i} \cos \theta_i + \sum_i \boldsymbol{F}_i^{\textrm{ext}} \cdot \boldsymbol{\hat{e}}_x, \label{eq1:app:rigid_2D} \\
M \frac{d^2 Y_c }{dt^2} +  \frac{d Y_c }{dt} &= \frac{1}{N} \sum_{i} \sin \theta_i + \sum_i \boldsymbol{F}_i^{\textrm{ext}} \cdot \boldsymbol{\hat{e}}_y, \label{eq2:app:rigid_2D} \\
I \left( M \frac{d^2 \phi }{dt^2} +  \frac{d \phi }{dt} \right) &= \sum_{i} \boldsymbol{r}_i^{\perp} \cdot \boldsymbol{\hat{n}}_i + \sum_{i} \boldsymbol{r}_i^{\perp} \cdot \boldsymbol{F}^{\textrm{ext}}_i, \label{eq3:app:rigid_2D} \\
 \frac{d \theta_i }{dt} = \frac{1}{\tau_n} \biggl( \boldsymbol{\hat{n}}_i^{\perp} \cdot& \frac{d}{dt} \begin{pmatrix} X_c \\ Y_c \end{pmatrix} + \boldsymbol{r}_i \cdot \boldsymbol{\hat{n}}_i \frac{d\phi}{dt} \biggr) + \sqrt{\frac{2D}{\tau_n}} \xi_i, \label{eq4:app:rigid_2D}
\end{align}
\end{subequations}
and $I = \sum_i r_i^2$ is the moment of inertia. Here, the arena is defined by a circle of pinned particles ($200$ particles positioned along a circle of radius $R_{\textrm{arena}} \simeq 3 \times L_{\textrm{solid}}$, with diameter $0.8$). Each pinned particle interacts with the active particles of the active solid with a purely repulsive WCA potential:
\begin{equation}
\boldsymbol{F}_i^{ext} = \sum_{j \in \mathcal{B}} \boldsymbol{F}^{\textrm{WCA}}_{j \rightarrow i} ( r_{ij} ),
\end{equation}
where $\mathcal{B}$ indicates the collection of pinned particles, with $r_{ij} = |\boldsymbol{r}_i - \boldsymbol{r}_j|$, and:
\begin{equation}
 \boldsymbol{F}^{WCA}_{j \rightarrow i}(r_{ij}) =
 \begin{cases}
        24 \varepsilon \left[ 2 \left( \frac{\sigma}{r_{ij}} \right)^{12} - \left( \frac{\sigma}{r_{ij}} \right)^6 \right] \boldsymbol{e}_{ij}, & \text{if } r_{ij} < r_{\textrm{cut}}, \\
        0, & \text{if } r_{ij} \geq r_{\textrm{cut}},
    \end{cases}
\end{equation}
where $r_{\textrm{cut}} = 2^{1/6} \sigma$, and with $\varepsilon = 1$ (stiffness-scale of the interaction with the boundary) and $\sigma = 0.9$ (interaction length scale with the boundary: half the stiff particles' diameter plus half the inter-active-particles' distances).

We simulate Eqs. (\ref{eq:app:rigid_2D}) for $D = 0.2$ and $M = 1$, where initially the active solid is at rest in the center of the arena, with randomly oriented active forces (Fig. \ref{fig:S9}A). Results are shown in the main text in Fig. 4C, as well as in Fig. \ref{fig:S9}. We find that the rigid active solid spontaneously adopt distinct modes of locomotion: (i) translational modes with $|\boldsymbol{V}| \simeq 1$ and vanishing $\omega$, with the active forces mostly aligned in a given direction (Figs. \ref{fig:S9}F-top); and (ii) rotational modes with vanishing $|\boldsymbol{V}|$ and $\omega \simeq \sqrt{N/I}$ (the maximum rotation rate, as obtained in steady state from fully azimuthal active forces \cite{hernandez2024model}), with the active forces adopting a centered, vortex-like configuration (Figs. \ref{fig:S9}F-bottom). As in the experiments, we find that the active solid can switch between the different modes upon collision with the boundaries of the arena (Figs. \ref{fig:S9}B-C). Finally, the different modes of locomotion are best illustrated by representing the distribution of velocity and rotation rate in the $V_{\parallel}-V_{\perp}$ and $|V|-\omega$ planes (Figs. \ref{fig:S9}D-E). We find that in this simplified model of rigid active solid with a square-like shape, all translational modes are equally likely (Figs. \ref{fig:S9}E).
\subsection{Numerical integration} Simulations in the 1D track and in a 2D disk-shaped arena are obtained by integrating Eqs. (\ref{eq:app:rigid_1D}) and (\ref{eq:app:rigid_2D}), respectively, with an Euler-Maruyama method with a fixed timestep $\delta t = 10^{-2}$. In both cases, the initial condition corresponds to a solid at rest in the center of the track/arena, with randomly oriented active forces.


\begin{figure}[p]
\centering
\includegraphics[width=\textwidth]{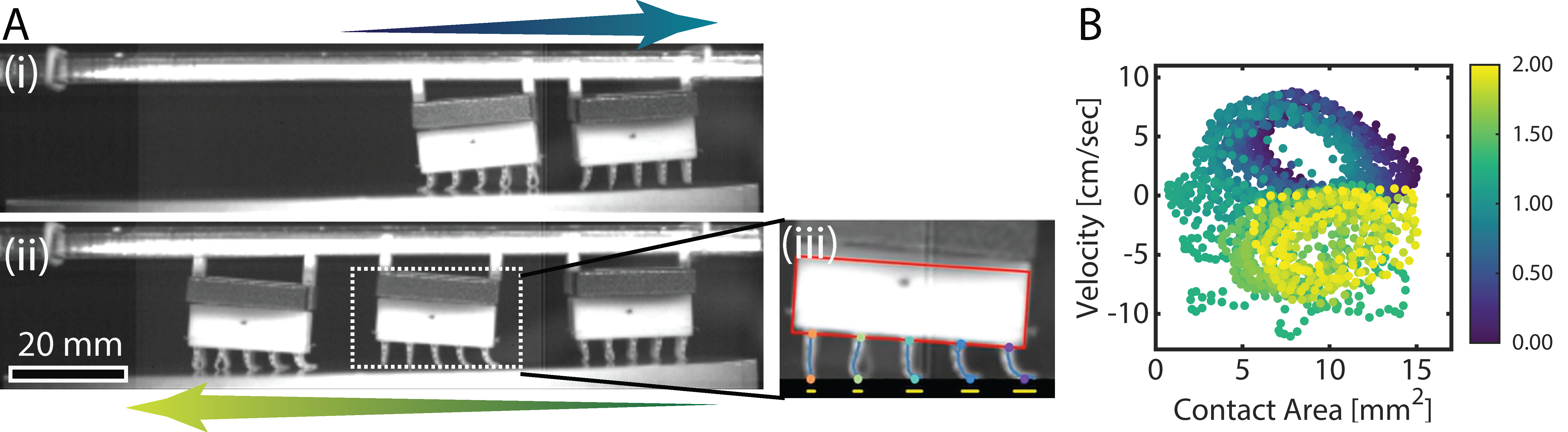}
\vspace{-0.1cm}
\caption{Locomotion in a 1D track in experiments: (A) The two modes of locomotion (left and right) are associated to (B) two distinct limit cycles in the plane of contact area $A$ vs. velocity $V_x$, where the contact area is obtained by segmentation of the cilia (see section \ref{sec:segmentation}). Collisions with the edges of the track can lead to a transition from one limit cycle to the other. The arrows in (i-ii) indicate the direction of motion and (iii) shows segmented contact points (in yellow) of the walker with the vibrating plate. The colorbar in B indicates time in seconds.}
\label{fig:S4}
\end{figure}

\begin{figure}[p]
\centering
\includegraphics[width=1.0\textwidth]{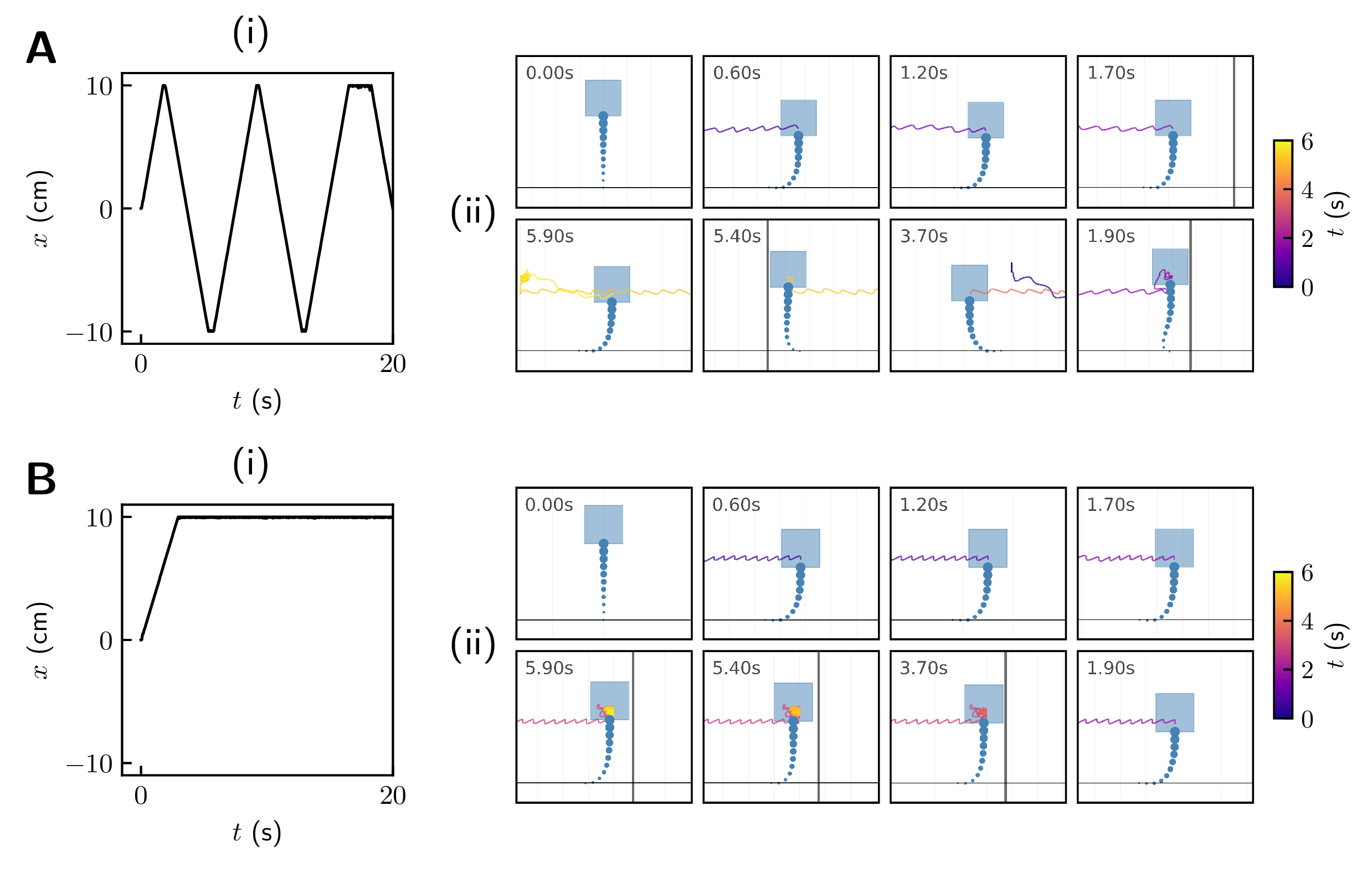}
\vspace{-0.7cm}
\caption{Locomotion in a 1D track in simulations of a single cilium (see section \ref{sec:single_cilium:dynamics}). (i) Representative trajectories in the 1D track of the top node of the cilium in the reversals (A) and trapped (B) regimes, for $\Gamma = 2.4$ and $\Gamma = 1.8$, respectively; arena size: $20$ cm; the cilium is initiated straight (with the bottom node just touching the surface), in the middle of the arena, and at rest. (ii) Simulation snapshots of the single cilium model illustrating the associated dynamics upon collisions; the trajectory of the top node of the cilium is overlaid and colored from blue to yellow as time increases. In all panels, $f = 40$ Hz.}
\label{fig:S5}
\end{figure}

\begin{figure}[p]
\centering
\includegraphics[width=0.8\textwidth]{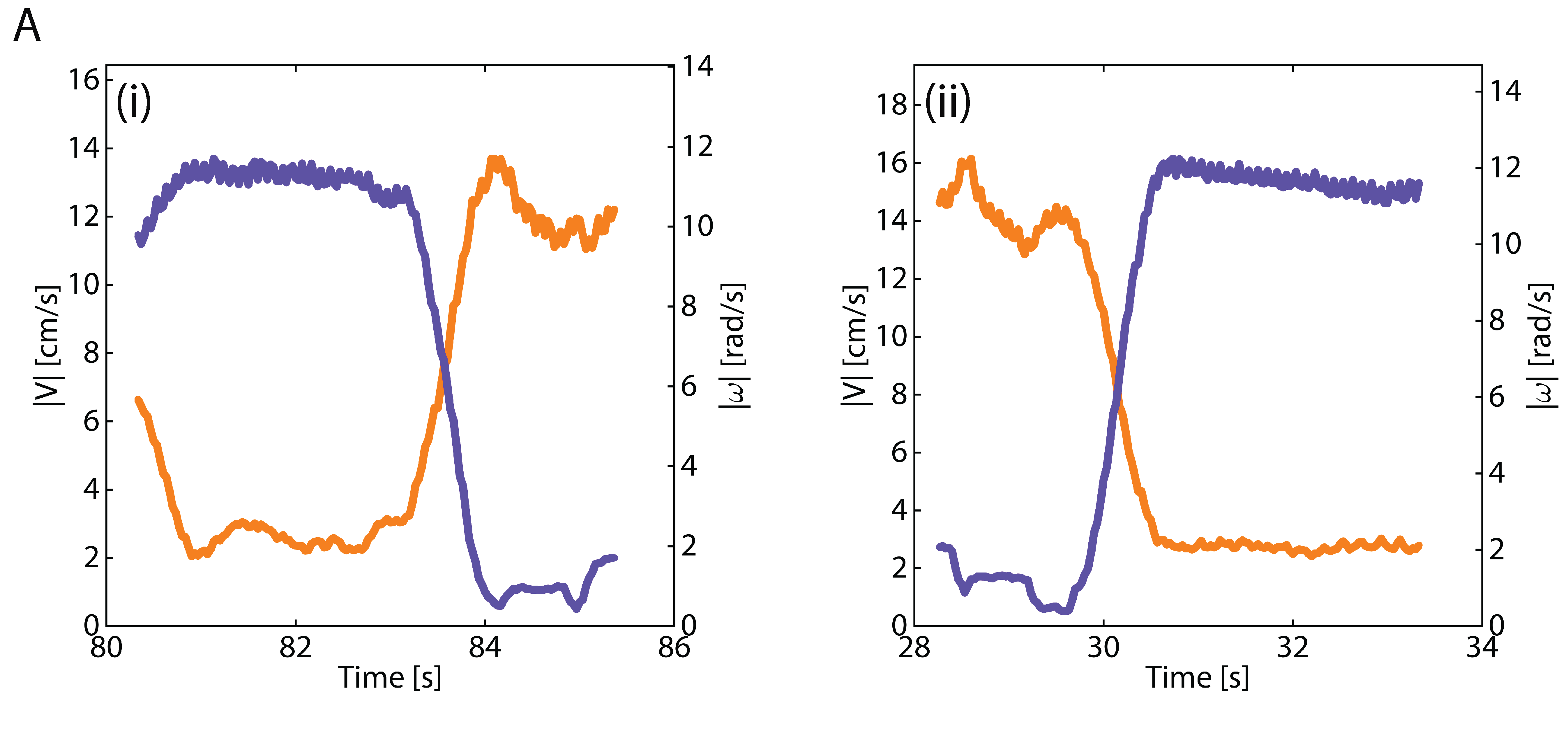}
\caption{Time evolution of the speed $|\boldsymbol{V}|$ (blue) and of the norm of the rotation rate $|\omega|$ (orange) during the collisions shown in Figs. 4A-iii,iv of the main text (as obtained from experiments). The panels illustrate transitions from the rotational to the translational regime (i) and from the translational to the rotational regime.}
\label{fig:S6}
\end{figure}

\begin{figure}[p]
\centering
\includegraphics[width=\textwidth]{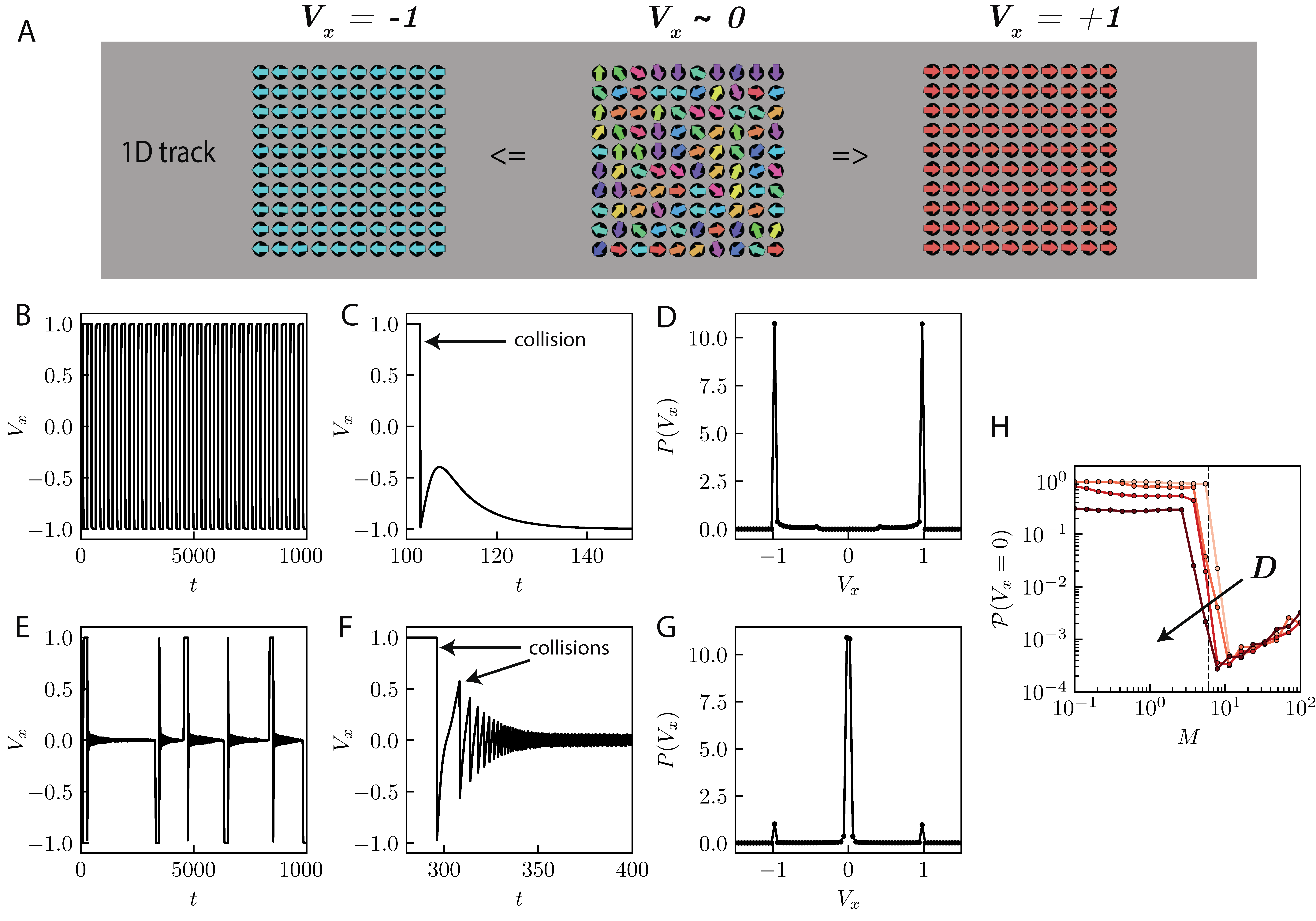}
\vspace*{-0.3cm}
\caption{Simulations of a rigid active solid in a 1D track, where the active units are arranged in a $10 \times 10$ square lattice (see section \ref{sec:active_solid:standard}). (A) Schematic of the simulations: at $t = 0$, the active solid is initiated with random orientations, at the center of the track, and at rest. After an initial transient, the active forces organize into one of two possible polarized configurations, and the solid translates to the left or to the right at constant velocity. The arrows represent the active forces and are colored according to their orientations. (B-G) Two distinct behaviors after a collision depending on the dimensionless inertia $M$, with a fixed, small noise $D = 10^{-3}$: (B-D) autonomous \textit{reversals} for large inertia $M \simeq 8$; (E-G) \textit{trapped} regime for small inertia $M \simeq 3$. (H) Statistical weight of the trapped state as a function of $M$ for various values of the noise $D \in \left[ 10^{-3}, 10^{-2.5}, 10^{-2.0}, 10^{-1.5} \right]$, color coded from light to dark red as $D$ increases; the vertical dashed line indicates $M = 6$.}
\label{fig:S7}
\end{figure}

\begin{figure}[p]
\centering
\begin{tikzpicture}
    
\fill[color=lightgray, opacity=0.5] (-1.8,2.5) rectangle (14.4,5.0);

\node[] at (6.3,3.75) {\includegraphics[height=2.5cm]{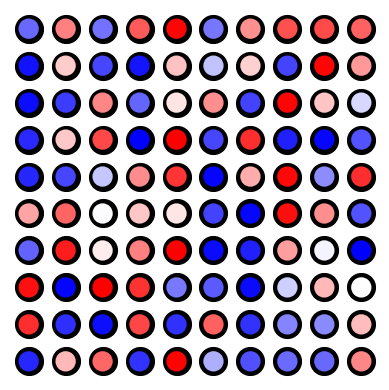}};

\node[] at (9.8,3.75) {\includegraphics[height=2.5cm]{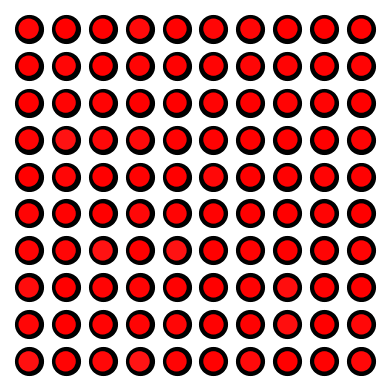}};

\node[] at (2.8,3.75) {\includegraphics[height=2.5cm]{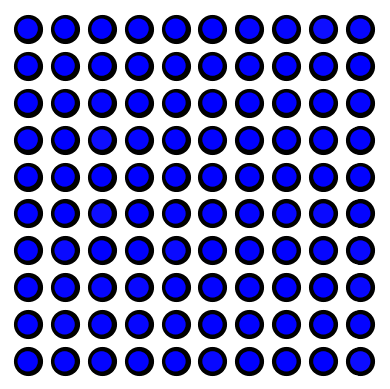}};

\draw[thick] (-1.8,2.5) rectangle (14.4,5.0);
\node[] at (-0.2,3.75) {\small \textbf{1D track}};

\node[] at (-1.5,5.3) {\small \textbf{A}};

\node[] at (6.3,5.3) {\small $V_x \simeq 0$};
\node[] at (9.8,5.3) {\small $V_x = +1$};
\node[] at (2.8,5.3) {\small $V_x = -1$};

\node[] at (8.05,3.75) {\small $\Rightarrow$};
\node[] at (4.55,3.75) {\small $\Leftarrow$};

\node[] at (-0.1,-0.1) {\includegraphics[height=4.1cm]{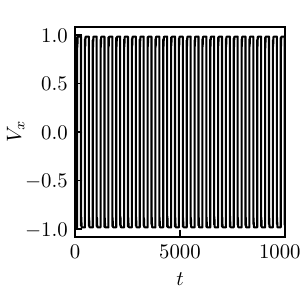}};
\node[] at (4.1,-0.1) {\includegraphics[height=4.1cm]{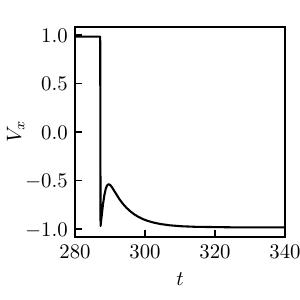}};
\node[] at (8.3,-0.1) {\includegraphics[height=4.1cm]{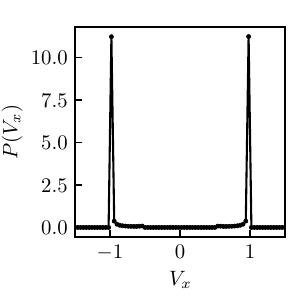}};

\node[] at (4.4,1.25) {\footnotesize collision};
\draw[->] (3.7,1.2) -- (3.5,1.2);

\node[] at (-1.7,1.9) {\small \textbf{B}};
\node[] at (2.5,1.9) {\small \textbf{C}};
\node[] at (6.7,1.9) {\small \textbf{D}};

\node[] at (-0.1,-4.4) {\includegraphics[height=4.1cm]{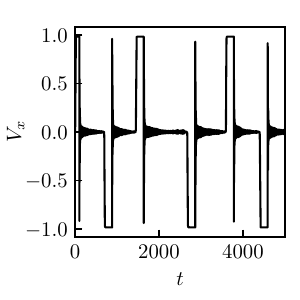}};
\node[] at (4.1,-4.4) {\includegraphics[height=4.1cm]{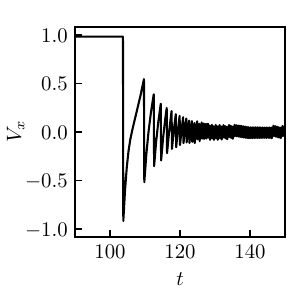}};
\node[] at (8.3,-4.4) {\includegraphics[height=4.1cm]{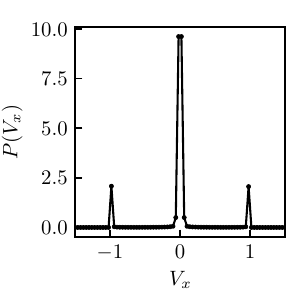}};

\node[] at (4.79,-3.03) {\footnotesize collisions};
\draw[->] (4.0,-3.08) -- (3.8,-3.08);
\draw[->] (4.15,-3.23) -- (4.03,-3.38);

\node[] at (-1.7,-2.4) {\small \textbf{E}};
\node[] at (2.5,-2.4) {\small \textbf{F}};
\node[] at (6.7,-2.4) {\small \textbf{G}};

\node[] at (12.7,-2.15) {\includegraphics[height=4.1cm]{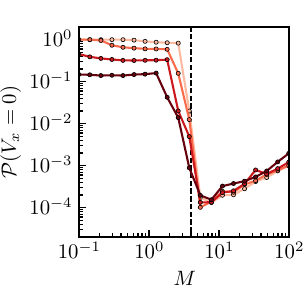}};

\draw[->] (13.05,-2.63) -- (13.75,-1.85);
\node[] at (14.06,-1.64) {\small $D \searrow$};

\node[] at (11.1,-0.15) {\small \textbf{H}};

\end{tikzpicture}

\vspace*{-0.3cm}
\caption{Simulations of a bidirectional rigid active solid in a 1D track, where the active units are arranged in a $10 \times 10$ square lattice (see section \ref{sec:active_solid:bidirectional}). (A) Schematic of the simulations: at $t = 0$, the active solid is initiated with random polarizations, at the center of the track, and at rest. After an initial transient, the active forces organize into one of two possible polarized configurations, and the solid translates to the left or to the right at constant velocity. The active units are colored from red to blue as the polarization $m_i$ goes from $-1$ to $1$. (B-G) Two distinct behaviors after a collision depending on the dimensionless inertia $M$, with a fixed, small noise $D = 10^{-3}$: (B-D) autonomous \textit{reversals} for large inertia $M \simeq 8$; (E-G) \textit{trapped} regime for small inertia $M \simeq 3$. (H) Statistical weight of the trapped state as a function of $M$ for various values of the noise $D \in \left[ 10^{-3}, 10^{-2.5}, 10^{-2.0}, 10^{-1.5} \right]$, color coded from light to dark red as $D$ increases; the vertical dashed line indicates $M = 4$.}
\label{fig:S8}
\end{figure}

\begin{figure}[p]
\centering
\begin{tikzpicture}
    \def\R{2.8}  
    \def\r{0.0373} 
    \def\N{200} 

    \fill[color=lightgray, opacity=0.5] (-9.0,-6.3) circle (\R);

    \foreach \i in {0,1,...,199} {
        \pgfmathsetmacro{\angle}{360/\N*\i}
        \pgfmathsetmacro{\x}{\R*cos(\angle)-9.0}
        \pgfmathsetmacro{\y}{\R*sin(\angle)-6.3}
        \fill[black] (
        \x,
        \y) circle (\r);
    }

\node[] at (-10.7,-3.6) {\small \textbf{A}};

\node[] at (-9.0,-6.3) {\includegraphics[height=0.933cm]{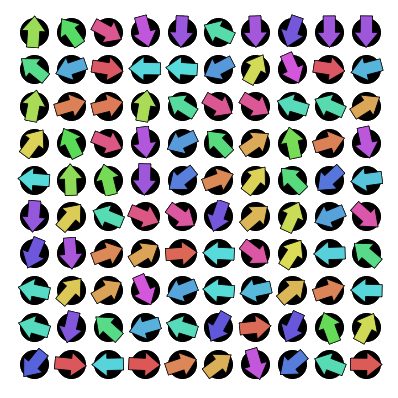}};
\node[] at (-9.0,-4.0) {\small \textbf{2D arena}};
\node[] at (-9.0,-7.0) {\small $t=0$};

\node[] at (0.95,-8.5) {\includegraphics[height=3.92cm]{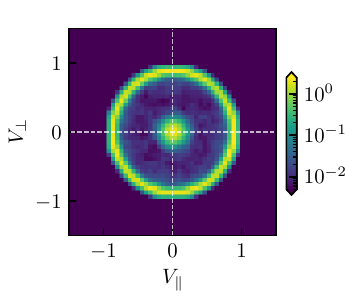}};
\node[] at (0.95,-4.4) {\includegraphics[height=3.92cm]{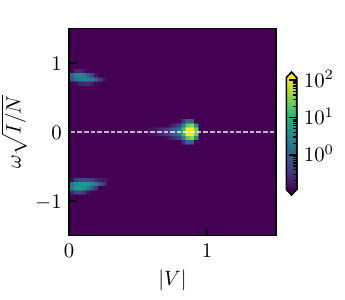}};

\node[] at (-3.95,-8.5) {\includegraphics[height=3.92cm]{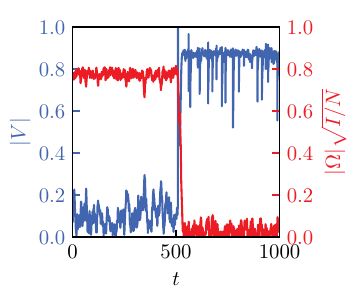}};
\node[] at (-3.95,-4.4) {\includegraphics[height=3.92cm]{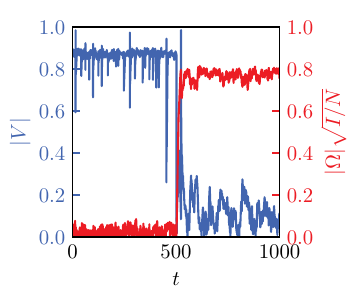}};

\node[rotate=-1.2] at (4.2,-5.4) {\includegraphics[height=1.8cm]{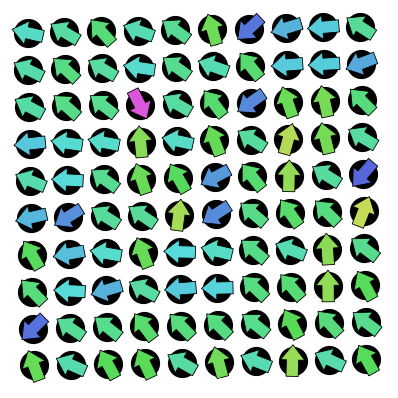}};
\node[] at (4.2,-7.4) {\includegraphics[height=1.8cm]{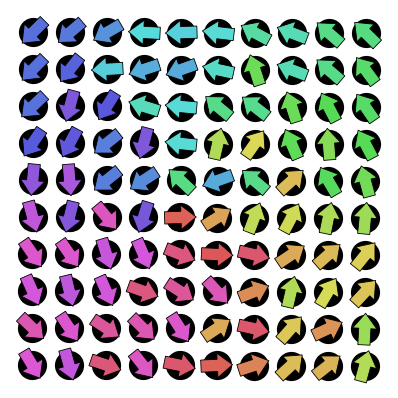}};

\draw[->, thick, color=black] (3.25,-5.6) -- (1.4,-4.3);
\draw[->, thick, color=black] (3.25,-5.6) -- (1.7,-7.6);

\draw[->, thick, color=black] (3.25,-6.8) -- (0.1,-5.0);
\draw[->, thick, color=black] (3.25,-6.8) -- (1.35,-8.2);

\node[] at (-5.7,-2.4) {\small \textbf{B}};
\node[] at (-5.7,-6.5) {\small \textbf{C}};

\node[] at (-0.9,-2.4) {\small \textbf{D}};
\node[] at (-0.9,-6.5) {\small \textbf{E}};

\node[] at (3.55,-4.2) {\small \textbf{F}};

\end{tikzpicture}

\vspace*{-0.3cm}
\caption{Simulations of a rigid active solid in a 2D disk-shaped arena, where the active units are arranged in a $10 \times 10$ square lattice (see section \ref{sec:active_solid:2d}). (A) Schematic of the simulations: at $t=0$, the active solid is initiated with random orientations, at the center of the arena, and at rest. The arrows represent the active forces and are colored according to their orientations. (B/C) Linear velocity and rotation rate as a function of time; collisions can trigger locomotion mode switching, e.g., from translational to rotational (B) or the opposite (C). (D/E) Probability distributions of the velocity and rotation rate, in the $|\boldsymbol{V}| - \omega$ plane (D), and in the $V_{\parallel} - V_{\perp}$ plane (E); the maps show the presence of both rotational and translational modes; the colormap is in log scale, as indicated. (F) Snapshots of the configuration of the active force field in the translational (top) and rotational (bottom) modes; same conventions as in (A). For all panels, $D = 0.2$, and $M = 1$.}
\label{fig:S9}
\end{figure}

\begin{figure}[p]
\centering
\begin{tikzpicture}

\node[] at (0.0,0.0) {\includegraphics[width=4.1cm]{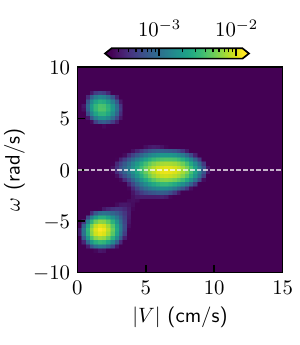}};

\node[] at (4.5,-0.2) {\includegraphics[width=4.0cm]{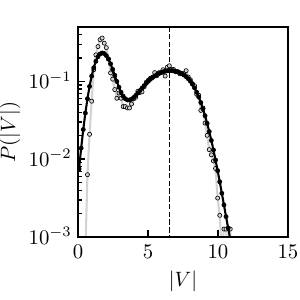}};

\node[] at (4.72,1.63) {\small $v_0$};

\node[] at (9.0,-0.2) {\includegraphics[width=4.0cm]{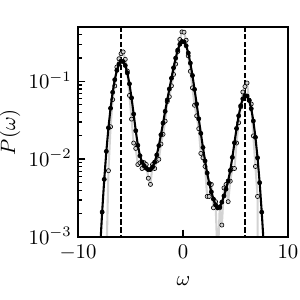}};

\node[] at (8.7,1.68) {\small $\omega_0^{-}$};
\node[] at (10.35,1.68) {\small $\omega_0^{+}$};

\node[] at (-1.7,1.55) {\small \textbf{A}};
\node[] at (2.8,1.55) {\small \textbf{B}};
\node[] at (7.3,1.55) {\small \textbf{C}};

\end{tikzpicture}

\vspace*{-0.3cm}
\caption{Determination of the cruise velocity $v_0$ and rotation rate $\omega_0$ (see section \ref{sec:peak_determination}). (A) Probability distribution of the velocity and rotation rate in the $|\boldsymbol{V}| - \omega$ plane, as obtained from experiments with isotropic ciliary walkers with fixed $f = 40$ Hz, $\Gamma = 2.45$, $W = 12$ g (same as in Fig. 4B of the main text); the maps show the presence of both rotational and translational modes; the colormap is in log scale. (B) Probability distribution of $|\boldsymbol{V}|$, as obtained from projecting (A) onto the $x$-axis. (C) Probability distribution of $\omega$, as obtained from projecting (A) onto the $y$-axis. The cruise velocity $v_0$ (resp. rotation rates $\omega_{0}^{\pm}$) is obtained from the position of the peak at large $|V|$ in the distribution of $|\boldsymbol{V}|$ (resp. from the positions of the peaks at finite $\omega$ in the distribution of $\omega$). The raw data are shown in gray; gaussian-filtered data are shown in black and are used for the peak detection.}
\label{fig:S10}
\end{figure}

\begin{figure}[p]
\centering
\includegraphics[width=\textwidth]{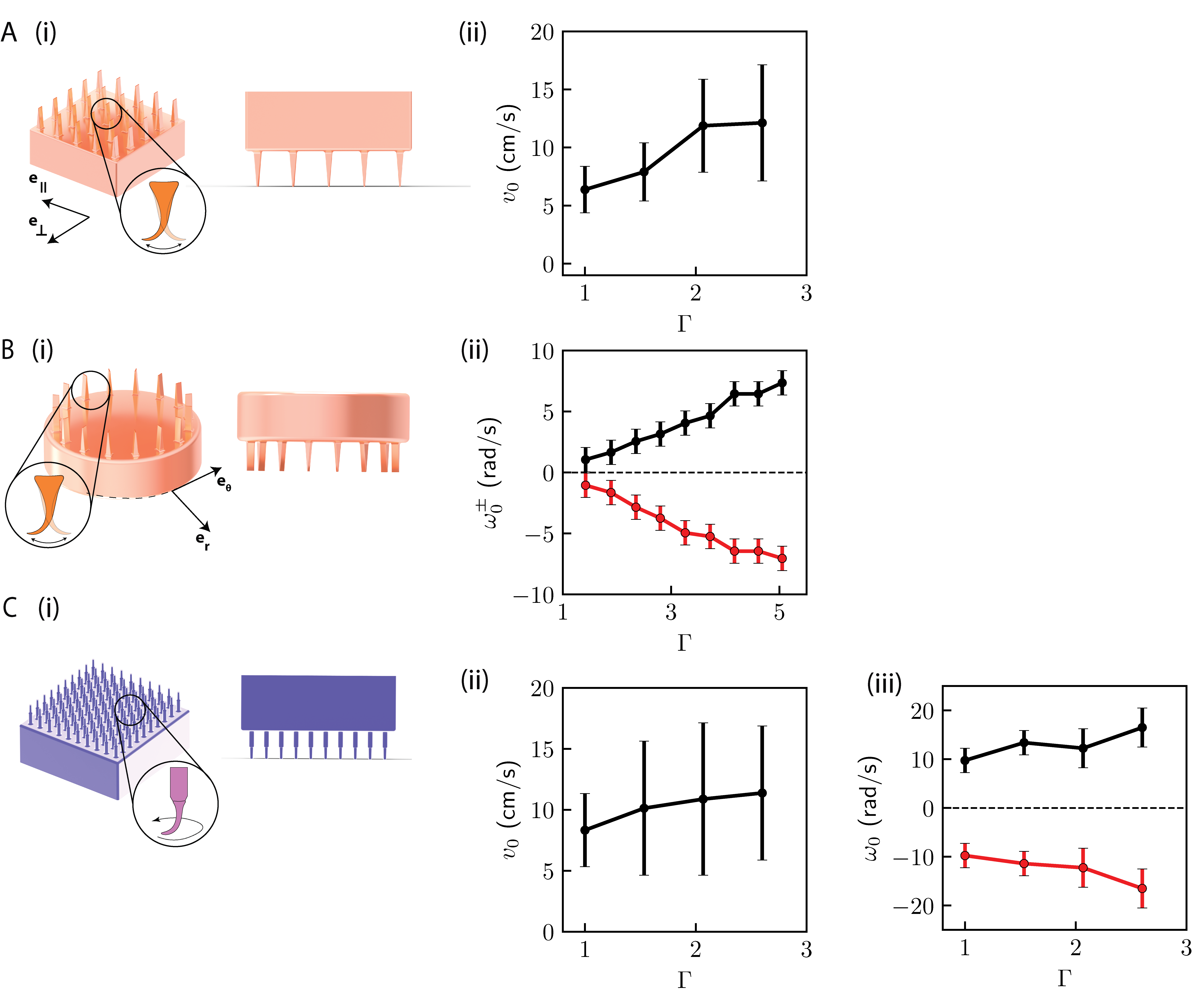}
\caption{Schematics of the different arrays of flexible cilia (i): (A) $5 \times 5$ square array of rectangular section cilia, all oriented along $\boldsymbol{\hat{e}}_{\parallel}$; 
(B) $16$ rectangular section cilia positioned along a ring, and all oriented along $e_{\theta}$; (C) $10 \times 10$ square array of cylindrical cilia. Different arrays of cilia give rise to drastically different locomotion behaviors: cruise velocity $v_{0}$ (A-ii, C-ii) and rotation rate $\omega_0$ (B-ii, C-iii), as obtained from the peaks in the experimental probability distributions (when they exist) as a function of the vibration amplitude $\Gamma$. The errorbars represent the spread of the probability distributions around the peaks, as obtained from the full width at half maximum (relative to nearby minima).}
\label{fig:S11}
\end{figure}

\newpage

\textbf{Movie S1}. Introduction to bidirectional walkers ($N$=25): (i) handheld motion of a bidirectional walker over a glass plate to illustrate the bistability of the rectangular-section buckled cilia, especially the two main buckled configurations of the cilia array; (ii) motion of a bidirectional walker on the vertically vibrated plate, confined in a disk-shaped arena ($W = 12$ g, $\Gamma = 1.97$, $f = 40$ Hz). \\

\textbf{Movie S2}. Different locomotion regimes of a bidirectional walker confined in a 1D track: (i) at low vibration amplitude $\Gamma$, the walker gets trapped at boundaries ($\Gamma = 2.93$, $W=18$ g, $f = 40$ Hz); (ii) at larger amplitudes $\Gamma$, the walker reverses its direction at boundaries ($\Gamma = 2.45$, $W=12$ g, $f = 40$ Hz) and (ii) shows erratic motion ($\Gamma = 2.93$, $W=8$ g, f = 40 Hz). All movies are recorded with a high-speed camera at $1000$ fps, and the playback speed is $200$ fps. \\

\textbf{Movie S3}. High-speed recordings show walkers with segmented contact area of the cilia with the vibrated plate, revealing that the bimodal locomotion ($\Gamma = 2.45$, $W=12$ g, $f = 40$ Hz) originates from two distinct limit cycles in response to the vertical vibration (see Fig. \ref{fig:S4}A-iii). \\

\textbf{Movie S4}. Simulation videos of a single cilium illustrating the trapped and reversals regimes in 1D track (see section \ref{sec:single_cilium}): (i) trapped regime ($\Gamma = 1.8$); (ii) reversing regime ($\Gamma = 2.4$); in  both cases $f = 40$ Hz, $W = 8.2$ g. \\

\textbf{Movie S5}. Introduction to isotropic walkers ($N=25$): (i) handheld motion of an isotropic walker over a glass plate to illustrate the multistability of the buckled cilia array; (ii) motion of an isotropic walker on the vertically vibrated plate, confined in a disk-shaped arena ($W = 12$ g, $\Gamma = 1.97$, $f = 40$ Hz). \\

\textbf{Movie S6}. Switching between locomotion modes of an isotropic ciliary walker ($N=25$): transitions (i) from translational to spinning and (ii) from spinning to translational modes of locomotion ($W = 12$ g, $\Gamma = 5.05$, $f = 60$ Hz). These transitions are also shown in Fig. 4A of the main text. \\

\textbf{Movie S7}. The rigid active solid model (see section \ref{sec:active_solid}) reproduces the behavior of isotropic walkers ($N=10 \times 10$): transition (i) from translational to spinning and (ii) from spinning to translational modes of locomotion. The arrows represent the active forces and are color coded according to their orientation (see inset colorbar in Fig. 4C of the main text). \\

\textbf{Movie S8}. Different design variants of ciliary walkers. Top view videos showing the motion of (i) bidirectional walkers ($W = 12$ g, $\Gamma = 2.45$, $f = 40$ Hz) with rectangular section cilia ($N=25$); (ii) birotational walkers ($\Gamma$ = 2.9, $f = 60$ Hz) with rectangular section cilia ($N=16$); and (iii) isotropic walkers ($\Gamma = 1.49$, $f = 40$ Hz) with cylindrical cilia ($N=100$). All the walkers are confined within in a disk-shaped arena. \\

\textbf{Movie S9}. Behavior of multiple isotropic walkers with cylindrical cilia ($N=100$): (i) showcasing behavior and tracking algorithm for $n = 1$, $4$ and $10$ walkers ($\Gamma = 1.5$, $f = 40$ Hz); (ii) removing one walker at a time starting with $n = 10$ walkers. \\

\end{document}